\documentclass[journal]{IEEEtran}
\usepackage{setspace}
\usepackage{amsfonts}
\usepackage[cmex10]{amsmath}
\usepackage{amsthm}

\usepackage{diagbox}   
\usepackage{multirow}
\usepackage{array}
\theoremstyle{definition}

\theoremstyle{remark}

\theoremstyle{remark}

\usepackage{pifont} 
\usepackage{array}
\usepackage{mdwmath}
\usepackage{mdwtab}
\usepackage{eqparbox}
\usepackage[caption=1]{caption}
\usepackage{subcaption}
\usepackage{url}
\usepackage{amssymb}
\usepackage{float}
\usepackage{indentfirst}
\usepackage{makeidx}
\usepackage{tabularx}
\usepackage{cases}
\usepackage{algorithmic}
\usepackage{graphicx}
\usepackage{subcaption} 
\usepackage{mathtools}
\usepackage{color}
\usepackage{bm}
\usepackage{stfloats}
\usepackage{lipsum}
\usepackage{amsfonts}
\usepackage{bm}
\usepackage{dsfont}
\usepackage{lipsum}
\usepackage{graphicx}
\usepackage{url}
\usepackage{bbm}
\usepackage{booktabs}
\usepackage{multirow} 
\usepackage{makecell} 
\newcommand{\RNum}[1]{\uppercase\expandafter{\romannumeral #1\relax}}
\ifCLASSOPTIONcompsoc
\usepackage[caption=false, font=normalsize, labelfont=sf, textfont=sf]{subfig}
\else
\usepackage[caption=false, font=footnotesize]{subfig}
\fi
\usepackage{multirow}
\allowdisplaybreaks
\makeatletter

\usepackage[linesnumbered,ruled,vlined]{algorithm2e}
\usepackage{amsmath}
\makeatother

\usepackage{amsthm}
\theoremstyle{definition} 
\theoremstyle{remark} 

\begin{document}
\title{MIMO-OFDM AI Receiver Based on Incrementally Conditioned Diffusion with Soft Decision}
\author{
    Weijie~Zhou,~
    Zhaoyang~Zhang,~
    Zhixian~Kong,~
  and~Zhaohui~Yang
\thanks{W. Zhou, Z. Zhang (\textit{Corresponding Author}), Z. Kong, and Z. Yang are with 1) College of Information Science and Electronic Engineering, Zhejiang University, Hangzhou 310027, China, and 2) Zhejiang Key Laboratory of Multi-modal Commu. Netw. \& Intell. Info. Proc., Hangzhou 310027, China (e-mails: \{wj\_zhou, ning\_ming, zhixian\_kong, yang\_zhaohui\}@zju.edu.cn). }
}
\maketitle
\begin{abstract}

Conventional iterative receiver usually begins with channel estimation using sparse pilot observations and follows with data detection based on the channel estimates, and then updates channel estimation using decision feedback, and so on. In Artificial Intelligence (AI)-based receiver design, it is also crucial to make use of such progressively enriched data observations to enhance the generative channel estimation. However, the statistical characteristics and reliability of the data decisions always evolve with the channel estimation processes, which brings great challenges to the design of the overall learning framework and algorithms. In this paper, we propose \textit{Diff-Rx}, an incrementally conditioned diffusion-based receiver with co-designed data detection and channel estimation, for multiple-input multiple-output orthogonal frequency division multiplexing (MIMO-OFDM) systems. Specifically, we develop a condition-adaptive post-training method which enables the generative channel estimator to adapt to the conditioning inputs that are progressively enriched by soft data decisions, and also to implicitly align the pilot- and data-induced channel feature spaces to mitigate potential estimation errors. The threshold-free soft decisions provide smooth condition updates without condition-specific reliability tuning. We further develop a conditional diffusion transformer that is capable of performing robust channel estimation under noisy observations and various pilot patterns while reducing the conventional multi-step diffusion generation to one single step. Simulations on both statistical and site-specific ray-tracing channels show that, Diff-Rx exhibits consistent gains across different noise levels, pilot densities and modulation schemes, and works well at a pilot density as low as 1/32 while achieving significant improvement in channel estimation and data detection performances.


\end{abstract}

\section{Introduction}

As sixth-generation wireless systems evolve towards the deep integration of communication, computing, sensing, and artificial intelligence (AI), AI-based multiple-input multiple-output orthogonal frequency division multiplexing (MIMO-OFDM) receivers are becoming an important component of the intelligence of the physical layer. In practical receivers, channel estimation and data detection are intrinsically coupled. Accurate channel state information (CSI) improves equalization and symbol detection, while detected data can be fed back to provide additional observations for subsequent channel refinement. Decision feedback is therefore an important mechanism to improve the performance of the receiver. A central challenge in AI receiver design is to jointly exploit pilot observations, received signals, and uncertain data-symbol information under limited pilot overhead, while maintaining low inference latency, robustness to various operating conditions, and an interpretable receiver processing chain.

Reliable channel estimation provides the foundation for such joint receiver processing. Conventional pilot-based methods recover the channel from the limited and noisy pilot observations. Least-squares (LS) estimation degrades under sparse pilots or low signal-to-noise ratios (SNRs), whereas linear minimum mean-square error (LMMSE) estimation requires channel covariance information and matrix inversion. Compressed sensing-based methods exploit channel sparsity and perform channel estimation using algorithms such as LASSO, orthogonal matching pursuit, and message passing (MP) \cite{6674179,7094443,7458188}, but depend on predefined sparsity assumptions and analytical channel models. Neural network (NN)-based estimators have also provided new solutions for receiver design. Convolutional \cite{8944280}, fully connected \cite{10445518}, transformer \cite{9526282}, and generative architectures \cite{9252921} can instead learn channel structures directly from data. Auxiliary information such as historical channel samples can also be incorporated to improve estimation performance and robustness \cite{10845822}. 

Diffusion models (DMs) have recently demonstrated strong capabilities in learning complex data distributions through progressive denoising \cite{song2020denoising}. For wireless data, the channel matrix exhibits structured correlations, while the received observations provide constraints on the generated channel \cite{meng2024diffusionmodelbasedposterior}. Channel estimation can therefore be formulated as a conditional generation problem, in which the receiver reconstructs a complete channel that is consistent with both the learned channel prior and the available observations. This formulation naturally allows pilot information, noise levels, received signals, and side information from data observations to be incorporated into a unified generative framework. Most DM-based channel estimators follow diffusion posterior sampling (DPS) \cite{chung2022diffusion}, which combines a learned channel prior with receiver observations during inference \cite{9957135,10930691}. Conditional DMs that directly incorporate pilot positions, user locations, or other side information further improve adaptability to various observation conditions \cite{mohsin2025conditionalpriorbasednonstationarychannel,yang2025diffusionmodelswirelesstransceivers,11142589}. These methods demonstrate the potential of DMs as channel estimation modules for AI receivers, but they mainly focus on pilot-conditioned channel estimation.

To alleviate the tradeoff between pilot overhead and receiver performance, classical iterative receivers introduce decision feedback into the joint channel estimation and data detection process. Methods based on expectation maximization (EM), MP, and iterative detection and decoding use the current channel estimate to infer transmitted data, and then feed hard decisions, posterior probabilities, or other information back as additional observations for channel re-estimation \cite{zhao2008iterative,park2015iterative,park2017expectation}. By alternating between channel estimation and data detection, these receivers progressively refine both tasks. Conventional methods usually rely on explicit selection strategies for the feedback, such as maximum-likelihood decisions, reliability thresholds, or mean-square error-based criteria, where only symbols considered sufficiently reliable are used as virtual pilots \cite{park2015iterative,park2017expectation}. However, when the initial channel estimate is inaccurate or the noise level is high, erroneous symbols may still be selected and propagated to subsequent iterations, resulting in error accumulation. Moreover, these selection mechanisms are generally local and driven by rules, making it difficult to fully exploit the global correlations among received signals, channel structures, pilot observations, and feedback information.

The principle of joint processing and decision feedback has also motivated AI receiver design. Among them, one representative direction is model-driven iterative receivers, while another is end-to-end neural receivers. Model-driven methods embed trainable components into conventional iterative procedures and learn dictionaries, filters, or message-update rules \cite{deka2026comprehensive,yang2025deep}. \cite{yang2025hybrid} further integrates NNs with belief propagation. These methods preserve explicit interactions between receiver modules and provide interpretability, but remain constrained by analytical models and update rules. End-to-end neural receivers instead jointly perform channel estimation, equalization, and detection and directly output log-likelihood ratios \cite{honkala2021deeprx,cammerer2023neural}. Although such receivers can optimize multiple processing modules jointly, their black-box nature makes the intermediate receiver states difficult to interpret and may require larger models with more parameters and higher computational complexity. Therefore, designing an efficient, adaptive, and interpretable AI receiver that tightly couples channel estimation with data detection remains an open problem.

Applying DMs within such an iterative AI receiver introduces further challenges. Most DM-based channel estimators rely on many denoising steps and additional observation-consistency corrections during inference. When channel generation is repeatedly invoked inside a decision-feedback loop, this cost is unbearable for practical receivers. Moreover, balancing the learned prior and observation constraints often requires manually selected scaling coefficients \cite{chung2022diffusion,10930691} that are sensitive to the SNR and pilot density. Some DM-based receivers further perform joint channel estimation and data detection \cite{zilberstein2023solvinglinearinverseproblems,cai2025jointactivitydetectionchannel,yang2025generativediffusionreceiversachieving}, where detected data may be reused as virtual pilots. However, feedback generation is often separated from the channel estimation network, which is not explicitly optimized for the resulting data observations. Irregular feedback may therefore deviate from the distributions of observations during training, while its uncertainty and relationship with the original observations cannot be fully learned. This weakens the mutual enhancement between channel estimation and data detection and may cause error propagation \cite{yang2025generativediffusionreceiversachieving}. Therefore, an effective DM-based iterative AI receiver must address both feedback utilization and the latency accumulated by repeated diffusion generation.

To address these challenges, we propose Diff-Rx, an incrementally conditioned DM-based AI receiver for MIMO-OFDM systems. Diff-Rx integrates data posterior inference, condition update, and channel estimation into a unified iterative procedure. The initial condition contains only pilot observations. Given the channel estimate, Diff-Rx evaluates the posterior distribution of each data symbol over the constellation set and constructs pseudo-channel observations using soft decision. These observations are combined with the pilots to update the condition, which is progressively enriched as its data-derived components evolve with the intermediate estimates. To effectively utilize such incremental feedback condition, we further post-train the channel estimator within the iterative receiver loop. The estimator thereby learns to handle the evolving data-aided observations and their global correlations with the original pilots. By learning from these in-loop conditions, the post-trained estimator implicitly bridges the distribution gap between pilot and data observations while mitigating the interference induced by channel and symbol estimation errors. The framework is also compatible with different update methods for transmitted data detection, including hard decisions, demonstrating its robustness and generality. By using soft decision, the proposed method avoids complicated threshold adjustments for selecting reliable data symbols under different conditions.

DMs are adopted for the generation of the channel owing to their strong conditional generation capability and noise robustness. Different from classical inverse problem DMs such as DPS, the proposed method does not rely on observation-consistency corrections during inference. Instead, the model is trained to directly learn the channel distribution conditioned on specific observations. Moreover, the proposed DM framework directly learns the target channels during training as the target prediction is more stable than noise prediction for high-dimensional data under limited network capacity \cite{li2026back}. The proposed DM network can recover the channel in one generation step, significantly reducing the inference latency of DM-based receivers. This is more suitable for low-latency AI-native MIMO-OFDM receiver deployment. The main contributions of this paper are summarized as follows:

\begin{itemize}

    \item We propose Diff-Rx, an incrementally conditioned DM-based receiver for MIMO-OFDM systems. It alternates between channel estimation and data posterior inference, progressively transitioning from pilot-only conditions to data-aided conditions through threshold-free soft decision.

    \item We develop a condition-adaptive post-training strategy that adapts a channel estimator to the varying data-derived observations in different iterations. The estimator learns to exploit the evolving uncertainty of data-derived observations and their correlations with pilot observations.

    \item We develop a conditional diffusion transformer (CDiT) as the channel estimation module of Diff-Rx to support different pilot patterns and noise levels. By directly predicting the target channel conditioned on receiver observations, CDiT avoids iterative observation-consistency corrections and supports one-step conditional generation. 

    \item Extensive simulations are conducted to evaluate channel estimation and data detection performance. The results demonstrate the consistent superiority of Diff-Rx under different settings.


\end{itemize}

\section{Preliminaries}
In this section, we briefly introduce the basic principles of the denoising diffusion probabilistic model (DDPM) \cite{NEURIPS2020_4c5bcfec}. 

DMs are one of the most powerful generative models. They learn the distribution of the data in the training set and, during inference, generate high-quality outputs by progressively removing noise from a random sample \cite{NEURIPS2020_4c5bcfec}. In the forward process, DMs progressively add Gaussian noise to the data $\textbf{H}_0 \sim q(\textbf{H}_0)$ in $T$ time steps according to a variance schedule $\beta_1, \dots ,\beta_T$ and finally transforms $\textbf{H}_0$ to white Gaussian noise. Each step in the forward process is given by
\begin{equation}
    q(\textbf{H}_t|\textbf{H}_{t-1}) = \mathcal{N}(\textbf{H}_t;\sqrt{1-\beta_t}\textbf{H}_{t-1},\beta_t\textbf{I}),
    \label{eq:DM1}
\end{equation}
where sample $\textbf{H}_t$ represents the latent variable at time step $t$, and $0<\beta_1<\dots<\beta_T<1$. The forward process admits sampling $\textbf{H}_t$ at an arbitrary time step $t$ in closed form
\begin{equation}
    q(\textbf{H}_t|\textbf{H}_0)=\mathcal{N}(\textbf{H}_t;\sqrt{\bar{\alpha}_t}\textbf{H}_0,(1-\bar{\alpha}_t)\mathbf{I}).\label{eq:DM2}
\end{equation}

For the reverse process, the DMs aim to generate the outputs starting at $p(\textbf{H}_{T})=\mathcal{N}(\textbf{H}_{T};\textbf{0},\textbf{I})$. However, due to the unknown distribution of $q(\textbf{H}_0)$, the conditional distribution $q(\textbf{H}_{t-1}|\textbf{H}_t)$ is typically intractable. Therefore the DMs model the reverse process as a Markov chain through a variational approach with learned Gaussian transitions  
\begin{equation}
    p_{\boldsymbol{\theta}}(\textbf{H}_{t-1}|\textbf{H}_t)=\mathcal{N}(\textbf{H}_{t-1};\boldsymbol{\mu}_{\boldsymbol{\theta}}(\textbf{H}_t,t),\boldsymbol{\Sigma}_{\boldsymbol{\theta}}(\textbf{H}_t,t)),
    \label{eq:DM3}
\end{equation}
where $\boldsymbol{\mu}_{\boldsymbol{\theta}}(\textbf{H}_t,t)$ and $\boldsymbol{\Sigma}_{\boldsymbol{\theta}}(\textbf{H}_t,t)$ are parameterized by  $\boldsymbol{\theta}$. Training is performed by optimizing the usual variational bound on negative log likelihood
\begin{equation}
    \begin{aligned}
&\mathbb{E}\left[-\log p_{\boldsymbol{\theta}}(\textbf{H}_{0})\right]\leq \mathbb{E}_{q}\left[-\log\frac{p_{\boldsymbol{\theta}}(\textbf{H}_{0:T})}{q(\textbf{H}_{1:T}|\textbf{H}_{0})}\right]\\
&=\mathbb{E}_{q}\left[\underbrace{D_{\mathrm{KL}}(q(\textbf{H}_{T}|\textbf{H}_{0})\parallel p(\textbf{H}_{T}))}_{L_{T}}\underbrace{-\log p_{\boldsymbol{\theta}}(\textbf{H}_{0}|\textbf{H}_{1})}_{L_{0}}\right.\\
& \quad+\left.\sum_{t>1}\underbrace{D_{\mathrm{KL}}(q(\textbf{H}_{t-1}|\textbf{H}_{t},\textbf{H}_{0})\parallel p_{\boldsymbol{\theta}}(\textbf{H}_{t-1}|\textbf{H}_{t}))}_{L_{t-1}}\right].
    \label{eq:KL}
\end{aligned}
\end{equation}

The term $L_{t-1}$ trains the parameters in (\ref{eq:DM3}) to perform one reverse diffusion step, which is tractable when conditioned on $\textbf{H}_0$ because $q(\textbf{H}_{t-1}|\textbf{H}_{t},\textbf{H}_{0})$ is also Gaussian
\begin{equation}
\begin{aligned}
    q(\textbf{H}_{t-1}|\textbf{H}_{t},\textbf{H}_{0})&=\mathcal{N}(\textbf{H}_{t-1};\tilde{\boldsymbol{\mu}}_{t},\tilde{\beta}_{t}\mathbf{I}),\\
    \tilde{\boldsymbol{\mu}}_{t}&=\frac{1}{\sqrt{\alpha_{t}}}\left(\textbf{H}_{t}-\frac{1-\alpha_{t}}{\sqrt{1-\bar{\alpha}_{t}}}\boldsymbol{\epsilon}_t\right),\\
    \tilde{\beta}_{t}&=\frac{1-\bar{\alpha}_{t-1}}{1-\bar{\alpha}_{t}}\beta_{t},
    \label{eq:DM4}
\end{aligned}
\end{equation}
where $\boldsymbol{\epsilon}_t\sim\mathcal{N}(\boldsymbol{0},\textbf{I})$ is the Gaussian noise in the time step $t$, $\alpha_t\overset{\Delta}{\operatorname*{=}}1-\beta_t$ and $\bar{\alpha}_t\overset{\Delta}{\operatorname*{=}}\prod_{i=1}^t\alpha_i$.

Different from the noise-prediction parameterization, we adopt the $x$-prediction method. Specifically, instead of predicting the Gaussian noise $\boldsymbol{\epsilon}_t$, the neural network directly estimates the channel $\mathbf{H}_0$ from its noisy version $\mathbf{H}_t$, where 
\begin{equation}
    \mathbf{H}_t=\sqrt{\bar{\alpha}_t}\mathbf{H}_0+\sqrt{1-\bar{\alpha}_t}\boldsymbol{\epsilon},\quad \boldsymbol{\epsilon}\sim\mathcal{N}(\mathbf{0},\mathbf{I}).
    \label{eq:Ht}
\end{equation}
According to the closed-form posterior in (\ref{eq:DM4}), the posterior mean can be equivalently written as
\begin{equation}
\begin{aligned}
\tilde{\boldsymbol{\mu}}_{t}
=
\frac{\sqrt{\bar{\alpha}_{t-1}}\beta_t}{1-\bar{\alpha}_t}\mathbf{H}_0
+
\frac{\sqrt{\alpha_t}(1-\bar{\alpha}_{t-1})}{1-\bar{\alpha}_t}\mathbf{H}_t .
\end{aligned}
\end{equation}
Therefore, under the $x$-prediction parameterization, the learned reverse mean is obtained by replacing the unknown $\mathbf{H}_0$ with the network prediction ${\mathbf{G}}_{\boldsymbol{\theta}}(\mathbf{H}_t,t)$ as
\begin{equation}
\begin{aligned}
\boldsymbol{\mu}_{\boldsymbol{\theta}}(\mathbf{H}_t,t)
=
\frac{\sqrt{\bar{\alpha}_{t-1}}\beta_t}{1-\bar{\alpha}_t}
{\mathbf{G}}_{\boldsymbol{\theta}}(\mathbf{H}_t,t)
+
\frac{\sqrt{\alpha_t}(1-\bar{\alpha}_{t-1})}{1-\bar{\alpha}_t}
\mathbf{H}_t .
\end{aligned}
\end{equation}
The corresponding training objective is defined directly in the clean-channel space as
\begin{equation}
\mathcal{L}_{\mathrm{DM}}(\boldsymbol{\theta})
=
\mathbb{E}_{\mathbf{H}_0,\boldsymbol{\epsilon},t}
\left[
\left\|
\mathbf{H}_0
-
{\mathbf{G}}_{\boldsymbol{\theta}}
\left(
\sqrt{\bar{\alpha}_t}\mathbf{H}_0
+
\sqrt{1-\bar{\alpha}_t}\boldsymbol{\epsilon},
t
\right)
\right\|_2^2
\right].
\label{eq:x_training}
\end{equation}

During inference, the reverse sampling step is performed as
\begin{equation}
\mathbf{H}_{t-1}
=
\boldsymbol{\mu}_{\boldsymbol{\theta}}(\mathbf{H}_t,t)
+
\sqrt{\tilde{\beta}_t}\boldsymbol{\epsilon}
\label{eq:ddpm_inference}
\end{equation}
iteratively from $t=T$ to $t=1$.

\begin{figure}
    \centering
    \includegraphics[width=1\linewidth]{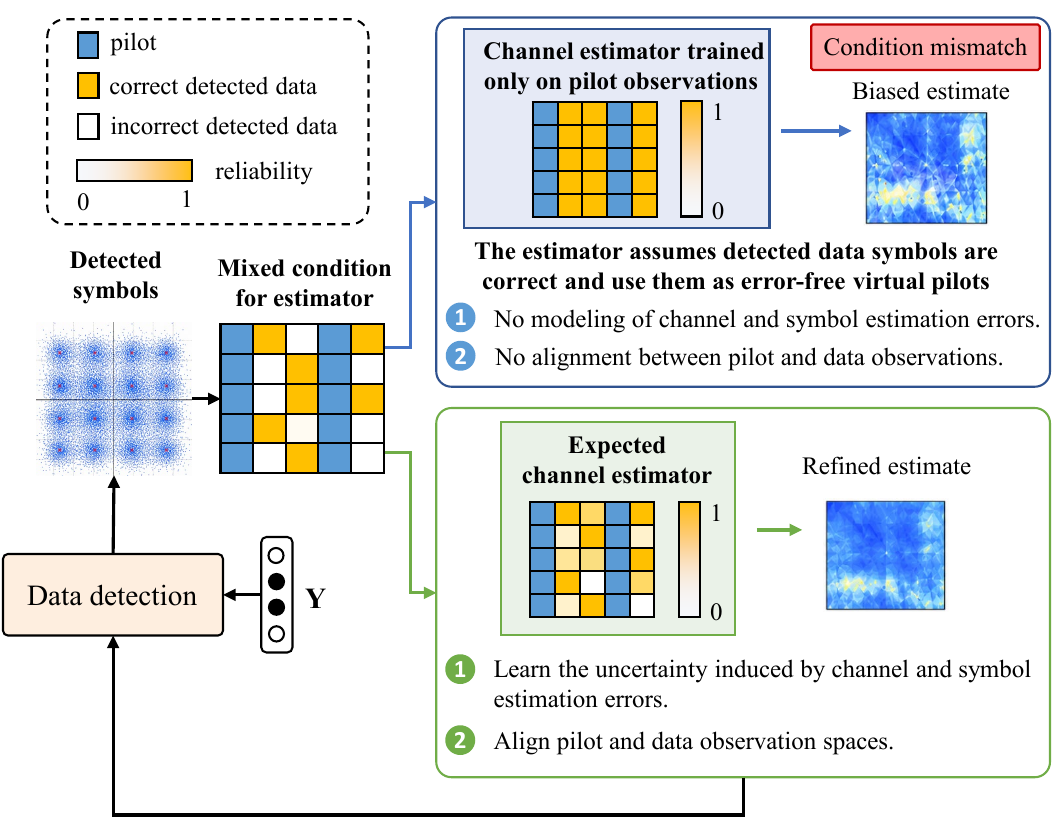}
    \caption{
    Conceptual comparison of decision-feedback utilization. Directly reusing a channel estimator trained only on pilot observations implicitly assumes that detected data symbols are correct and treats them as error-free virtual pilots. Since channel and symbol estimation errors are not modeled and pilot and data observations are not aligned, erroneous feedback may cause biased channel estimation. The expected estimator should mitigate estimation errors and aligns the two observation spaces for channel refinement.}
    \label{fig:design motivation}
    \vspace{-0.3cm}
\end{figure}

\section{System Model}
In this paper, we consider an OFDM uplink system with $N_\mathrm{f}$ subcarriers. There is a base station (BS) equipped with $N_\mathrm{r} \gg 1$ antennas serving multiple single-antenna user equipments (UEs). We denote the channel frequency response (CFR) as $\textbf{H}\in \mathbb{C}^{N_\mathrm{r}\times N_\mathrm{f}}$. The transmitted symbol vector, denoted by $\textbf{x}\in \mathbb{C}^{N_\mathrm{f}}$, is carried on the $N_\mathrm{f}$ subcarriers. Some subcarriers are selected for pilot symbols and others are for data symbols. Here, we denote the sets of the indices corresponding to the pilot and the data symbols as $\boldsymbol{\mathcal{P}}$ and $\boldsymbol{\mathcal{D}}$, respectively. The pilot symbols $\mathbf{x}_i, i\in \boldsymbol{\mathcal{P}}$, are known by the receiver. Meanwhile, the data symbols $\mathbf{x}_i, i\in \boldsymbol{\mathcal{D}}$, are chosen from a specific constellation. The data and pilot symbols would be selected from different constellations. We assume that the constellation points have a uniform prior distribution. Thus, the received signals denoted by $\textbf{Y}\in\mathbb{C}^{N_\mathrm{r}\times N_\mathrm{f}}$ for the specific symbol $x_i, i\in\boldsymbol{\mathcal{P}}\cup \boldsymbol{\mathcal{D}}$ can be expressed as
\begin{equation}
    \textbf{y}_i = x_i\textbf{h}_i + \textbf{n}_i,
    \label{eq:channel model1}
\end{equation}
where $\textbf{y}_i$ represents the $i$-th column of $\textbf{Y}$ corresponding to the $i$-th subcarrier, $\textbf{h}_i$ represents the $i$-th column of $\textbf{H}$ and $\textbf{n}_i\sim \mathcal{CN}(0, \sigma_{\mathrm{n}}^2\textbf{I})$ is Gaussian noise. $\sigma_{\mathrm{n}}^2$ is the noise variance. The equation (\ref{eq:channel model1}) can be written as 
\begin{equation}
    \textbf{Y}=\textbf{H}\textbf{X}+\textbf{N}, \quad \textbf{X} = \mathrm{diag}\{\textbf{x}\},
    \label{eq:channel model2}
\end{equation}
where \textbf{N} is the Gaussian noise.
Based on the known pilot symbols, the initial channel response at pilot positions  $\widetilde{\textbf{h}}_i,i\in\boldsymbol{\mathcal{P}}$, can be computed as $ \widetilde{\textbf{h}}_i = {\textbf{y}_i}/{x_i}=\textbf{h}_i+\widetilde{\textbf{n}}_i$, where $\widetilde{\textbf{n}}_i={\textbf{n}_i}/{x_i}\sim \mathcal{CN}(0, \sigma_{\mathrm{n}}^2 / |x_i|^2\textbf{I})$. 

\begin{figure*}
    \centering
    \includegraphics[width=1\linewidth]{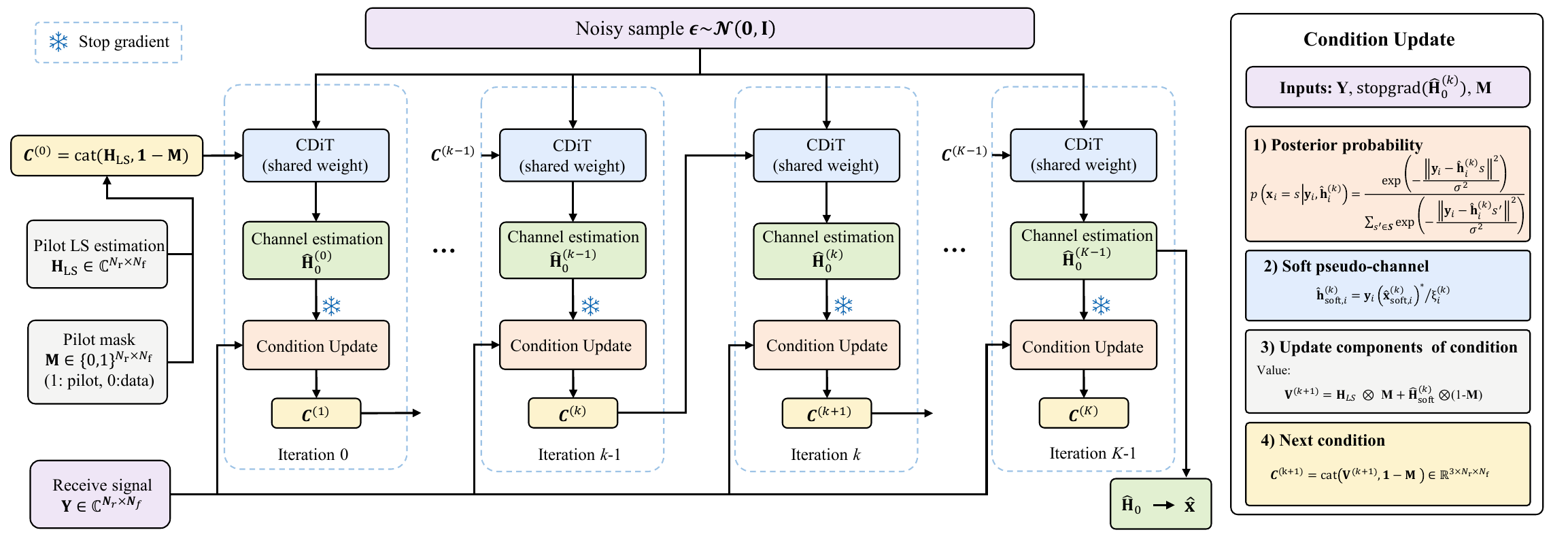}
    \caption{The illustration of the Diff-Rx framework, where the pretrained CDiT progressively refines the channel estimation using incremental conditions updated with pseudo-channel information.}    
    \label{fig:unfold-framework}
    \vspace{-0.5cm}
\end{figure*}

Therefore, we aim to recover $\textbf{H}$ and data symbols based on the observed noisy channel on partial subcarriers, which corresponds to the raw estimates $\widetilde{\textbf{h}}_i$ at the pilot positions, as well as the received signals $\textbf{Y}$. The problem can be modeled as 
\begin{equation}
\begin{aligned}
         \mathrm{g}_0: \left [\widetilde{\textbf{h}}_i \right ]_{i\in\boldsymbol{\mathcal{P}}}  \in \mathbb{C}^{N_\mathrm{r} \times\left | \boldsymbol{\mathcal{P}}\right| }, \textbf{Y}  &\to \textbf{H}\in \mathbb{C}^{N_\mathrm{r} \times N_\mathrm{f}}, \mathbf{x},\\
\end{aligned}
    \label{eq:origin channel model}
\end{equation}
where $\left|\boldsymbol{\mathcal{P}}\right|$ represents the number of elements of $\boldsymbol{\mathcal{P}}$. 

The pilot positions in a single observation correspond to a specific pilot pattern, which is known by the receiver. We represent the pilot pattern using the mask $\textbf{M}\in\{0,1\}^{N_\mathrm{r}\times N_\mathrm{f}}$ where 1 indicates the positions of pilot symbols and 0 indicates the positions of data symbols. $\textbf{M}$ is a striped matrix with all ones in $i$-th columns where $i\in\boldsymbol{\mathcal{P}}$ and zeros elsewhere. The channel estimation problem in (\ref{eq:origin channel model}) can be transformed to the following mathematical model:  
\begin{equation}
\begin{aligned}
         \mathrm{g}: \widetilde{\textbf{H}} \in \mathbb{C}^{N_\mathrm{r}\times N_\mathrm{f}}, \textbf{Y}  &\to \textbf{H}\in \mathbb{C}^{N_\mathrm{r}\times N_\mathrm{f}}, \mathbf{x}_i, i\in\boldsymbol{\mathcal{D}},\\
     \widetilde{\textbf{H}} &= (\textbf{H} + \widetilde{\textbf{N}})\otimes\textbf{M},
\end{aligned}
\label{eq:channel model3}
\end{equation}
where $\otimes$ represents the element-wise multiplication and $\widetilde{\textbf{N}}$ is the concatenation of $\widetilde{\textbf{n}}_i$. The the noise variance $\widetilde{\textbf{n}}_i, i\in\boldsymbol{\mathcal{D}}$ at non-pilot positions do not affect our solution after multiplying the CFR matrix by $\textbf{M}$. Thus, for simplicity, $\widetilde{\textbf{N}}$ is a unified representation for the noise.

\section{Proposed Methods}

\subsection{Motivations on the Framework Design}


Decision feedback can improve channel estimation by introducing additional observations at data positions. However, as shown in Fig.~\ref{fig:design motivation}, directly applying an estimator trained only on pilot observations to such data-aided inputs may cause a substantial distribution mismatch. Such an estimator observes only noisy channel estimates obtained from the exactly known pilot symbols. In contrast, observations at data positions are constructed from symbols inferred using the estimated channel and are affected by both receiver noise and estimation errors. Their reliability also varies with the SNR and modulation order. The estimator may treat pilot-derived and data-derived observations as equally reliable, causing erroneous feedback to bias subsequent channel estimation and data detection. Moreover, data-derived observations are unavailable during the initial estimation stage and are different as the receiver updates its symbol estimates. Therefore, an effective iterative receiver should incrementally incorporate newly inferred data information while accounting for the distinct and evolving uncertainty of the added observations.

Another motivation for the proposed architecture is to avoid directly learning channel estimation and data detection within a single neural network. The transmitted data lie in a high-dimensional discrete space, making direct neural modeling difficult. In contrast, wireless channels exhibit structured correlations and can often be represented by a low-dimensional manifold. Therefore, we use a classical data detection method to efficiently transform data domain information into channel domain, and then employ the neural network to learn and refine the structured channel representation. In this section, we will introduce the proposed framework in detail.

\subsection{The Proposed Diff-Rx Framework}
We propose a Diff-Rx framework, which integrates posterior symbol inference into a CDiT-based diffusion channel estimator. The illustration of the framework is shown in Fig.~\ref{fig:unfold-framework}. 
\subsubsection{Iterative channel estimation in Diff-Rx}

Diff-Rx is a trainable iterative AI receiver inspired by the alternating estimation principle of conventional EM-like receivers. At each receiver iteration, Diff-Rx first estimates the channel conditioned on the pilot observations and the currently available data-symbol information. The resulting channel estimate is then used to update the posterior distribution of the data symbols and reconstruct the condition for the next iteration. The same noise sample for diffusion is shared across all receiver iterations. Consequently, the difference between two consecutive channel estimates mainly results from the updated conditioning information rather than from different realizations of diffusion noise. This design enables Diff-Rx to progressively refine the channel estimation through a controlled and 
consistent iterative process.

Diff-Rx performs a maximum of $K$ channel estimation iterations. Specifically, the first iteration essentially corresponds to channel estimation using an estimator trained only on pilot observations. In this stage, the condition is constructed by concatenating $\textbf{H}_{\mathrm{LS}}$, with the reversed pilot mask $\textbf{1}-\textbf{M}$, and is used to guide CDiT to generate the initial channel estimate. At the $k$-th iteration, CDiT estimates the channel $\widehat{\mathbf{H}}_0^{(k)}$ using the condition $\mathbf{C}^{(k)}$ provided by the previous iteration. Unlike the first iteration, $\mathbf{C}^{(k)}$ contains not only $\textbf{H}_{\mathrm{LS}}$ and the mask information, but also the pseudo-channel estimate inferred from the data symbol in the ($k$-1)-th iteration. Therefore, the channel generation network can adaptively exploit pseudo-channel estimate from data symbols, which provides additional side information for refining the current channel estimation. 

After $K$ iterations, Diff-Rx produces the final channel estimate $\widehat{\textbf{H}}_0$. Moreover, considering the strict latency requirement of wireless communication systems, the proposed architecture naturally supports early termination, where an intermediate channel estimation can be directly used as the output.

\subsubsection{Condition Update}
The condition update is performed at each iteration. At the $k$-th iteration, Diff-Rx reconstructs the next condition $\mathbf{C}^{(k+1)}$ using the current channel estimate $\widehat{\mathbf{H}}_0^{(k)}$, the received signal $\mathbf{Y}$, and the pilot mask $\mathbf{M}$. Instead of explicitly preserving the data symbol selection results from the $(k-1)$-th iteration, Diff-Rx recomputes the posterior distribution and soft decision based on $\widehat{\mathbf{H}}_0^{(k)}$. This avoids directly reusing early decisions and mitigates their propagation across receiver iterations.

For simplicity, we consider the single-stream case. Given $\widehat{\mathbf{H}}_0^{(k)}$, $\mathbf{Y}$, and $\mathbf{M}$, we approximate the posterior probability of each data symbol over the constellation set using the current channel estimate as
\begin{equation}
    p\left(\textbf{x}_i = s \middle| \mathbf{y}_i, \widehat{\mathbf{h}}_i^{(k)}\right) = \frac
{\exp\left(-{\left\|\mathbf{y}_i - \widehat{\mathbf{h}}_i^{(k)} s\right\|^2}/{\sigma_\mathrm{n}^2}\right)}
{\sum_{s' \in \mathcal{S}} \exp\left(-{\left\|\mathbf{y}_i - \widehat{\mathbf{h}}_i^{(k)} s'\right\|^2}/{\sigma_\mathrm{n}^2}\right)},
\label{eq:iterative:condition_update_post}
\end{equation}
where $\widehat{\mathbf{h}}_i^{(k)}$ is the $i$-th column of 
$\widehat{\mathbf{H}}_0^{(k)}$, $\boldsymbol{\mathcal{S}}$ represents the constellation generating the data symbols, $s\in\boldsymbol{\mathcal{S}}$ and $i\in\boldsymbol{\mathcal{D}}$. After obtaining the posterior probabilities of each data symbol over the constellation points, the soft pseudo-channel is obtained by minimizing the following reconstruction error as
\begin{equation}
\begin{aligned}
    \widehat{\mathbf{h} }_{\mathrm{soft},i}^{(k)} &= \arg\min_{\mathbf{\bar{h}}_i} \; \mathbb{E}\left[\left\|\mathbf{y}_i - \mathbf{x}_i \mathbf{\bar{h}}_i\right\|^2 \mid \mathbf{y}_i, \hat{\mathbf{h}}_i^{(k)}\right] \\
    &={\mathbf y_i\big(\widehat{\mathbf{x}}_{\mathrm{soft}, i}^{(k)}\big)^*}/{\xi_i^{(k)}},
\end{aligned}
\label{eq:iterative:H_soft}
\end{equation}
where 
\begin{equation}
        \widehat{\mathbf{x}}_{\mathrm{soft}, i}^{(k)} 
        =\mathbb E\left[\mathbf{x}_i\mid\mathbf y_i,\widehat{\mathbf h}_i^{(k)}\right] =\sum_{s'\in\mathcal{S}}p(\mathbf{x}_i=s' \mid \mathbf{y}_i,\widehat{\mathbf{h}}_i^{(k)})s'
        \label{eq:iterative:soft_symbol_estimate_vector}
\end{equation}
and 
\begin{equation}
    \quad    \xi_i^{(k)}\triangleq\mathbb E\left[|\mathbf{x}_i|^2\mid\mathbf y_i,\widehat{\mathbf h}_i^{(k)}\right].
\end{equation}

Such soft decision avoids prematurely discretizing uncertain data symbols. In low-SNR or high-order modulation scenarios, an erroneous hard decision can be directly propagated into the subsequent pseudo-channel update, leading to error accumulation across iterations. In contrast, the soft decision preserves the uncertainty information contained in the posterior distribution and allows data symbols to participate in the condition update in a smoother manner.

To characterize the uncertainty contained in the pseudo-channel observation, define the channel and soft symbol estimation errors as $\Delta\mathbf h_i^{(k)} \triangleq \mathbf h_i-\widehat{\mathbf h}_i^{(k)}$ and $\Delta \mathbf{x}_i^{(k)}\triangleq \mathbf{x}_i-\widehat{\mathbf{x}}_{\mathrm{soft},i}^{(k)},$ respectively. We further define $\nu_i^{(k)}\triangleq \xi_i^{(k)}-\left|\widehat{\mathbf{x}}_{\mathrm{soft}, i}^{(k)}\right|^2$. Thus the pseudo-channel observation can also be expressed as
\begin{equation}
    \widehat{\mathbf h}_{\mathrm{soft},i}^{(k)}
=
\mathbf h_i+\boldsymbol\eta_i^{(k)},
\end{equation}
where
\begin{equation}
\begin{aligned}
       \boldsymbol\eta_i^{(k)}
= &\frac{\big(\widehat{\mathbf{x}}_{\mathrm{soft},i}^{(k)}\big)^*}
{\xi_i^{(k)}}
\left(
\widehat{\mathbf h}_i^{(k)}\Delta \mathbf{x}_i^{(k)}
+
\Delta\mathbf h_i^{(k)}\mathbf{x}_i
+
\mathbf n_i
\right)\\
&-
\frac{\nu_i^{(k)}}{\xi_i^{(k)}}
\widehat{\mathbf h}_i^{(k)}
-
\Delta\mathbf h_i^{(k)}. 
\label{eq:residual_error}
\end{aligned}
\end{equation}

Therefore, the data-derived pseudo-channel consists of the true channel and an equivalent residual jointly induced by channel estimation error, symbol estimation error and receiver noise. A channel estimator trained only on pilot-derived observations is not optimized for this coupled residual. By post-training the estimator with the pseudo-channel conditions generated inside the iterative receiver, Diff-Rx learns to implicitly compensate for the residual in (\ref{eq:residual_error}), without explicitly estimating or cancelling its individual components. As the channel and symbol estimates are progressively refined, the residual can be progressively reduced, providing a more reliable condition for the subsequent iteration.

The complete pseudo-channel matrix $\widehat{\mathbf{H}}_{\text{soft}}^{(k)}$ is obtained by stacking $\widehat{\mathbf{h}}_{\text{soft},i}^{(k)}$ over the data subcarriers. During the condition update, this pseudo-channel is injected into the next-iteration condition as 
\begin{equation}
    \mathbf{V}^{(k+1)} = \mathbf{H}_\mathrm{LS} \otimes \mathbf{M} + \widehat{\mathbf{H}}_{\text{soft}}^{(k)} \otimes (\mathbf{1}-\mathbf{M}) .
    \label{eq:iterative:value_update}
\end{equation}
As a result, the data-derived pseudo-channel can provide additional pseudo-pilot information for channel estimation. 

Finally, we update the condition for the next iteration as 
\begin{equation}
        \mathbf{C}^{(k+1)}
          =
          \mathrm{cat}
          \left(
          \mathbf{V}^{(k+1)},
          \mathbf{1}-\mathbf{M}
          \right).
          \label{eq:iterative:next_cond}
\end{equation}
We maintain the mask $\mathbf{M}$ to emphasize the pilot positions.

\subsubsection{Data Detection}

After completing $K$ receiver iterations, Diff-Rx obtains the final 
channel estimate $\widehat{\mathbf{H}}_0=\widehat{\mathbf{H}}_0^{(K-1)}$. The posterior probability of each data symbol over the constellation set is recomputed according to \eqref{eq:iterative:condition_update_post}. For each data subcarrier 
$i\in\mathcal{D}$, the constellation point with the maximum posterior 
probability is selected as the final detection result:
\begin{equation}
\widehat{\mathbf{x}}_i
=
\underset{s\in\mathcal{S}}{\arg\max}
\;
p\left(
\mathbf{x}_i=s
\,\middle|\,
\mathbf{y}_i,\widehat{\mathbf{h}}_i
\right),
\qquad i\in\mathcal{D},
\label{eq:detection}
\end{equation}
where $\widehat{\mathbf{h}}_i$ is the $i$-th column of 
$\widehat{\mathbf{H}}_0$.

\begin{figure*}[t]
\centering
  \includegraphics[width=0.9\linewidth]{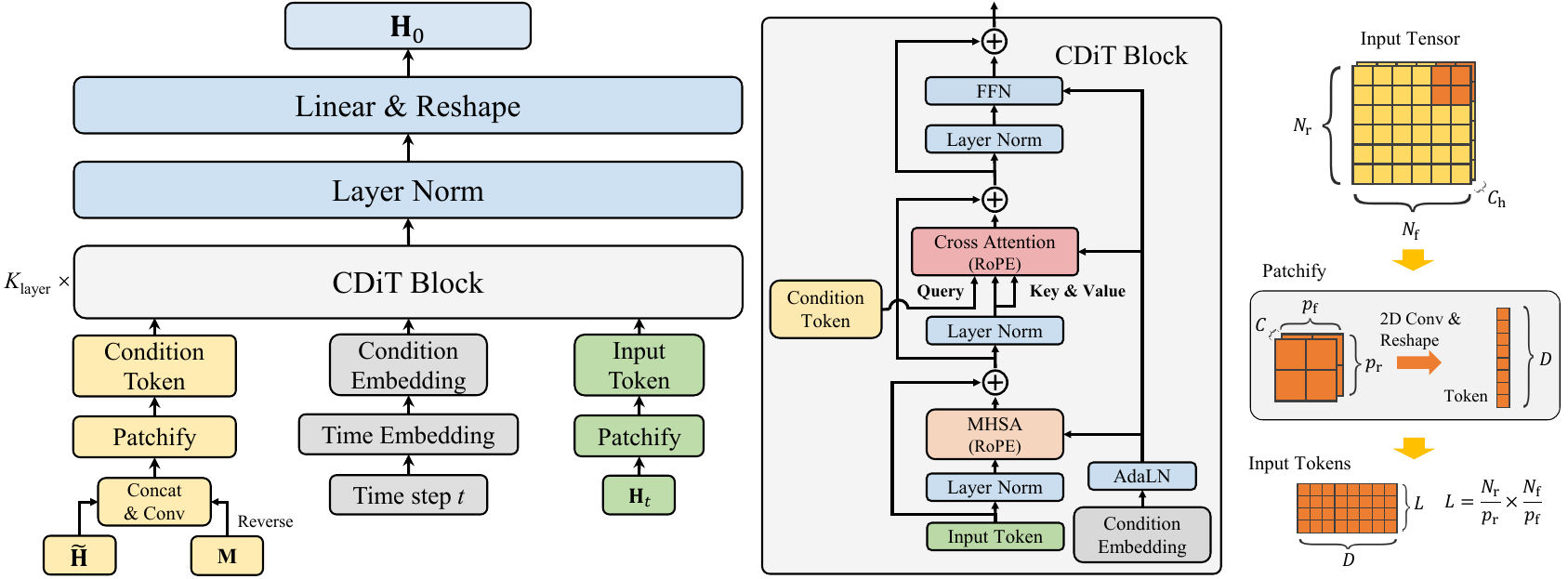}
  \caption{The proposed CDiT architecture. At timestep $t$, given the channel estimation at pilot positions and the mask $\mathbf{M}$, which indicates the position of the pilots, the network denoises $\mathbf{H}_t$ and estimates the original noiseless channel $\mathbf{H}_0$. } 
  \label{fig:CDiT model}
  \vspace{-0.5cm}
\end{figure*}
 
\subsection{The Proposed CDiT Channel Estimation Framework}

In practical systems, the receiver should handle various SNRs and pilot patterns, which motivates the use of DMs for robust conditional channel generation under diverse observations. Our proposed channel estimation model is shown in Fig. \ref{fig:CDiT model}. Given that CFR is complex-valued, we separate its real and imaginary components and stack them along a new dimension. Thus, both the noised channel matrices ${\textbf{H}}_t$ denoted in (\ref{eq:Ht}) at time step $t$ and the raw estimate $\widetilde{\textbf{H}}$ have the shape ${2\times N_\mathrm{r}\times N_\mathrm{f}}$. During the training phase, ${\textbf{H}}_t$ represents the ground-truth channel ${\textbf{H}}_0$ corrupted by noise at timestep $t$. During the inference phase, ${\textbf{H}}_t$ represents the intermediate noisy channel obtained during the denoising process. Both represent the channels corrupted by noise, thus we uniformly use ${\textbf{H}}_t$ to describe this variable.

\subsubsection{Patchify}
We implement patchify \cite{Peebles_2023_ICCV} via 2D convolution to learn intra-patch features between subcarriers and antennas. Assuming both the kernel size and the stride size are $p_{\mathrm{r}}\times p_{\mathrm{f}}$ and the embedding size of the tokens is $D$, we can transform an input tensor of dimensions ${C_\mathrm{h}\times N_\mathrm{r}\times N_\mathrm{f}}$ into $L \times D$, where the sequence length $L = (N_{\mathrm{r}}/p_{\mathrm{r}})\times (N_{\mathrm{f}}/p_{\mathrm{f}})$ and $C_\mathrm{h}$ is the number of channels. We perform different patchify modules for the generation of input tokens and the condition tokens because the dimensions of their inputs are different. The schematic illustration is shown in the right part of Fig. \ref{fig:CDiT model}. 

\subsubsection{Conditional embedding}
For scalar conditional information, we embed the diffusion timestep $t$ into the model. The scalars are encoded by sinusoidal embeddings followed by MLPs, and the outputs are added to form the global conditioning vectors. The global conditioning vector is injected into each CDiT block through adaptive layer normalization (AdaLN) \cite{Peebles_2023_ICCV}. Specifically, a modulation MLP maps condition embedding to shift, scale, and gate parameters, which are used to modulate the normalized features.

\subsubsection{CDiT block}

In DMs, the design of the denoising network also plays a critical role in the final recovery performance. This work adopts a DiT-based denoising network \cite{Peebles_2023_ICCV} to enhance the modeling of global correlations and improve channel recovery performance under sparse pilot conditions. Specifically, the noised channel tokens and condition tokens are fed into a sequence of CDiT blocks. We denote the number of CDiT blocks as $K_{\mathrm{layer}}$. Following the DiT model design, each transformer block includes a multi-head self-attention (MHSA) module and a point-wise feedforward (FFN) module. 

Instead of directly concatenating the conditional information at the input \cite{yan2025diffnmr3advancingnmrresolution, 9887996}, we apply the cross-attention module to incorporate the conditional input tokens. Here, the conditional input token serves as the query, while the output from the previous self-attention acts as both the key and the value. This design is driven by the need for deep condition-aware fusion. The cross-attention mechanism explicitly models the correlation between the conditions and the latent features of the noised channel, allowing the network to adaptively attend to the most relevant conditional information based on the current context. Furthermore, we apply AdaLN to the cross-attention module. We adopt rotary position encoding (RoPE) \cite{su2024roformer} to encode the relative positions of tokens.



\section{Training and Inference Method \label{section:improve speed}}

\subsection{Training Method}

\subsubsection{Data preprocessing for training of CDiT}
The amplitude of the CFR data exhibits significant variations. However, DMs typically require inputs to be normalized within the range of $[-1, 1]$ for better performance. Thus we normalize the data by dividing each sample $\mathbf{H}_0$ by a power factor which can be derived as $\sqrt{\|\textbf{H}_0\|_\mathrm{2}^2/({N}_\mathrm{r}{N}_\mathrm{f})}$. This ensures that the input amplitudes remain approximately on the order of 1, resulting in a more uniform sample distribution and effectively reducing the learning difficulty for the network. 

\subsubsection{Training of CDiT}

After the preprocessing described above, we can obtain a set of accurate channel samples $\mathbf{H}_0$. During training, we perform mixed training with noise of various variances and different pilot patterns. Specifically, before the sample is fed into the network, we randomly generate noise and pilot patterns. After incorporating conditional information $\{\widetilde{\mathbf{H}}, \mathbf{M} \}$, denoted as $\boldsymbol{C}$, into the training, we adopt a direct $x$-prediction objective. CDiT is trained to estimate $\mathbf{H}_0$ itself rather than the injected diffusion noise. The training objective of CDiT is expressed as
\begin{equation}
\mathcal{L}_{\mathrm{CDM}}(\boldsymbol{\theta})
=\mathbb{E}_{\mathbf{H}_0,\boldsymbol{\epsilon},t}
\left[
\left\|
\mathbf{H}_0-
\mathbf{G}_{\boldsymbol{\theta}}
\left(
\mathbf{H}_t,
t,
\boldsymbol{C}
\right)
\right\|_2^2
\right].
\end{equation}

In the proposed conditional DM framework, the final objective is to obtain an accurate channel estimation rather than to recover the artificial diffusion noise. This choice is motivated by the manifold perspective of structured high-dimensional data discussed in~\cite{li2026back}. The channel $\mathbf{H}_0$ exhibits physical correlations in the frequency and antenna domains and can be regarded as lying on a low-dimensional structured manifold embedded in the high-dimensional observation space. In contrast, the injected Gaussian noise $\boldsymbol{\epsilon}$ is unstructured and occupies the full high-dimensional space. Directly predicting clean data is more suitable in such settings because the network only needs to recover the structured data component rather than model the full-dimensional random noise. This makes the training loss more consistent with downstream channel estimation metrics and avoids indirectly recovering the channel through an additional reparameterization step after noise prediction. 


To enable CDiT to handle arbitrary pilot-pattern inputs, we introduce a random masking strategy during training. Specifically, let the pilot interval be $P$, and we randomly choose the starting pilot position $P_{\mathrm{start}}$ from the range $[0, P-1]$. The uniformly spaced pilot indices in the subcarrier domain are $P_{\mathrm{start}} + wP$, where $w=0,1,.., \lfloor{N_{\mathrm{f}}}/{P}\rfloor-1$.
Based on this index set, a basic comb-type pilot mask is constructed. We then randomly activate additional frequency positions by setting their mask values to one, thereby generating nonuniform random pilot patterns. With this training strategy, CDiT not only maintains its estimation performance under uniformly spaced pilot patterns, but also improves its adaptability to random pilot patterns and dynamic pseudo-pilot inputs.

The corresponding training procedure is summarized in Algorithm~\ref{alg:training}.

\subsubsection{Amplitude-aligned normalization}
Because both the input and the output channel of CDiT are normalized in training, it is important to align the amplitude scale during the training and inference pipeline of Diff-Rx. We estimate the channel normalization factor from the LS estimation because the true channel amplitude is unknown during inference. We first compute the power of $\widetilde{\mathbf{H}}$ in (\ref{eq:channel model3}) as
\begin{equation}
P_{\mathrm{LS}}
=
\frac{1}{N_{\mathrm{r}}|\boldsymbol{\mathcal{P}}|}
\sum_{i\in \boldsymbol{\mathcal{P}}}
\left\|
\widetilde{\mathbf{h}}_i
\right\|^2_2 .
\label{eq:inference:amp_norm}
\end{equation}
Since the LS estimate satisfies the $\widetilde{\mathbf{h}}_i=\mathbf{h}_i+\mathbf{n}_i/x_i$, the noise power is scaled by the known pilot symbol amplitude. Therefore, given the linear SNR $r$, the normalization amplitude is estimated as
\begin{equation}
\widehat{a}
=
\sqrt{
\frac{P_{\mathrm{LS}}}{1+
{P_{\mathrm{x}} \eta_{\mathrm p}}/{r}}
},
\quad
\eta_{\mathrm p}
=
\frac{1}{|\boldsymbol{\mathcal{P}}|}
\sum_{i\in \boldsymbol{\mathcal{P}}}
\frac{1}{|x_i|^2},
\label{eq:eudr_amp_estimation}
\end{equation}
where $P_{\mathrm{x}}=\mathbb{E}[||\mathbf{x}||^2]$ is the average symbol power. As the constellation is normalized, we approximate $P_X=1$. The initial LS condition, CDiT output, and pseudo-channel are all aligned using the same amplitude factor $\widehat{a}$.

{
\setlength{\textfloatsep}{2pt plus 1pt minus 2pt}
\begin{algorithm}[t!] 
  \caption{Training Algorithm for CDiT}
  \label{alg:training}
  \begin{algorithmic}[1]
    \REQUIRE SNR range $[r_{\mathrm{min}},r_{\mathrm{max}}]$
    \REPEAT
      \STATE Sample $\textbf{H}_0 \sim q(\textbf{H}_0)$ 
      \STATE Sample power $P_{\mathrm{signal}}=\|\textbf{H}_0\|_\mathrm{F}^2/({N}_\mathrm{r}{N}_\mathrm{f})$
      \STATE $\textbf{H}_0 = \textbf{H}_0/\sqrt{P_{\mathrm{signal}}}$
      \STATE Sample $t \sim \text{Uniform}(\{1, \dots, T\})$         
      \STATE Sample $\boldsymbol{\epsilon} \sim \mathcal{N}(0, \mathbf{I})$  
      \STATE Sample a random mask $\textbf{M}$. 
      \STATE Sample $r$ from $[r_{\mathrm{min}}, r_{\mathrm{max}}]$, $\sigma_{\mathrm{n}}^2 = 1/r$, sample $\textbf{N}\sim \mathcal{CN}(0, \sigma_{\mathrm{n}}^2\textbf{I})$
      \STATE $\widetilde{\textbf{H}}=(\textbf{H}_0+\textbf{N})\otimes\textbf{M}$
      \STATE $\boldsymbol{C}\leftarrow\{{\widetilde{\textbf{H}}}, \textbf{M}\}$
      \STATE Take gradient descent step on $\nabla_{\boldsymbol{\theta}} \left\|\textbf{H}_0 - \textbf{G}_{\boldsymbol{\theta}}(\sqrt{\bar{\alpha}_t}\textbf{H}_0 + \sqrt{1 - \bar{\alpha}_t}\boldsymbol{\epsilon}, t, \boldsymbol{C})\right\|^2$  
    \UNTIL converged                                                
  \end{algorithmic}
\end{algorithm}
}

\subsubsection{Training of Diff-Rx}

Different from the standard single-pass CDiT training, Diff-Rx refines the channel estimate over $K$ receiver iterations. During training, the same noisy sample $\mathbf{H}_t$ is shared by all iterations. The shared-weight CDiT estimator at iteration $k$ predicts the target channel as
\begin{equation}
\widehat{\mathbf{H}}_0^{(k)}
=
\mathbf{G}_{\boldsymbol{\theta}}
\left(
\mathbf{H}_t,
t,
\mathbf{C}^{(k)}
\right),
\quad k=0,\ldots,K-1 .
\label{eq:training:eudr_stage_prediction}
\end{equation}

To facilitate Diff-Rx training, we initialize CDiT using the pretrained model. The pretrained CDiT is trained for pilot-conditioned channel estimation and is independent of the modulation, providing a strong channel prior. This initialization yields reliable early-stage channel estimation, which reduces the risk of propagating severely biased pseudo-channel observations and stabilizes the iterative training process. Moreover, after the post-training of Diff-Rx, the network becomes adapted to the data-aided update process under the target constellation. Therefore, starting from a pretrained CDiT allows the proposed framework to adapt to different constellations with only limited post-training, instead of training a separate estimator from scratch for each modulation.

\begin{algorithm}[t!]
  \caption{Training Algorithm of Diff-Rx}
  \label{alg:entropy_unfolding_training}
  \begin{algorithmic}[1]
    \REQUIRE SNR range $[r_{\mathrm{min}},r_{\mathrm{max}}]$, pretrained CDiT $\mathbf{G}_{\boldsymbol{\theta}}$, maximum number of iterations $K$, constellation $\mathcal{S}$.
    \REPEAT
      \STATE Sample $\mathbf{H}_0 \sim q(\mathbf{H}_0)$, $\mathbf{M}\in\{0,1\}^{N_{\mathrm r}\times N_{\mathrm f}}$, $\mathbf{X}\in\mathcal{S}^{N_{\mathrm f}}$ and $r$ from $[r_{\mathrm{min}},r_{\mathrm{max}}]$.
      \STATE Generate received signal $\mathbf{Y}=\mathbf{H}_0\mathbf{X}+\mathbf{N}$ at $\mathrm{SNR}=r$.
      \STATE Obtain LS estimation
      $\mathbf{H}_{\mathrm{LS}}=\mathbf{Y}/\mathbf{X}$.
      \STATE Estimate the amplitude factor $\widehat{a}$ using (\ref{eq:inference:amp_norm}) and normalize  $\mathbf{H}_{\mathrm{LS}}\leftarrow\mathbf{H}_{\mathrm{LS}}/\widehat{a}$.
      \STATE Sample $t\sim\mathrm{Uniform}(\{1,\dots,T\})$ and $\boldsymbol{\epsilon}\sim\mathcal{N}(\mathbf{0},\mathbf{I})$.
      \STATE Construct the noisy sample $        \mathbf{H}_t=\sqrt{\bar{\alpha}_t}\mathbf{H}_0+\sqrt{1-\bar{\alpha}_t}\boldsymbol{\epsilon}$.
      \STATE Initialize $\mathbf{C}^{(0)}
        =
        \mathrm{concat}
        \left(
        \mathbf{H}_{\mathrm{LS}}\otimes\mathbf{M},
        \mathbf{1}-\mathbf{M}
        \right)$.
      \STATE Set $\mathcal{L}=0$.
      \FOR{$k=0$ to $K-1$}
        \STATE Estimate the channel $\widehat{\mathbf{H}}_0^{(k)}
          =
          \mathbf{G}_{\boldsymbol{\theta}}
          \left(
          \mathbf{H}_t,
          t,
          \mathbf{C}^{(k)}
          \right)$.
        \STATE $\mathcal{L}
          \leftarrow
          \mathcal{L}
          +
          \left\|
          \widehat{\mathbf{H}}_0^{(k)}
          -
          \mathbf{H}_0
          \right\|_2^2$.
        
        \STATE $\textbf{Detach}: \widehat{\mathbf{H}}_0^{(k)} \leftarrow \mathbf{stopgrad}(\widehat{\mathbf{H}}_0^{(k)}\times\widehat{a})$.

        \STATE Compute posterior probability according to (\ref{eq:iterative:condition_update_post}).
        \STATE Compute normalized pseudo-channel $\widehat{\mathbf{H}}_{\mathrm{soft}}^{(k)}/\widehat{a}$.
        \STATE Update $\mathbf{V}^{(k+1)}$ according to (\ref{eq:iterative:value_update}).
        \STATE Update the next condition $\mathbf{C}^{(k+1)}$ according to (\ref{eq:iterative:next_cond}).

      \ENDFOR
      \STATE Take gradient descent step on $\nabla_{\boldsymbol{\theta}} \frac{1}{K}\mathcal{L}$.
    \UNTIL converged
  \end{algorithmic}
\end{algorithm}

We stop the gradient through the condition update. Specifically, at the $k$-th iteration, the CDiT estimator first produces a channel estimate. Then we use a detached version of the estimate to compute the next condition as 
\begin{equation}
\mathbf{C}^{(k+1)}=\Phi\left(\mathbf{Y},\mathrm{sg}\left(\widehat{\mathbf{H}}_0^{(k)}\right),\mathbf{M}\right),
\label{eq:training:detach}
\end{equation}
where $\mathrm{sg}(\cdot)$ denotes the stop-gradient operation and $\Phi(\cdot)$ is the condition update module.  

This design is essential for stable training of Diff-Rx. Without stop-gradient, the condition update would be included in the differentiable recurrent computational graph, forcing gradients to propagate through posterior normalization, and pseudo-channel reconstruction across iterations. These operations form a long and numerically sensitive gradient path, which can lead to invalid gradient propagation and unstable optimization.

The stop-gradient operation also makes the training consistent with an EM-like alternating inference structure. The posterior inference acts as an E-step-like operation that generates data-dependent soft side information from the current channel estimation, while CDiT performs channel estimation under the generated condition. The posterior quantities produced by the inference step are treated as fixed when updating the estimator. Therefore, the model-based condition update is used only to generate fixed conditional inputs for the next iteration, rather than being optimized through backpropagation as a differentiable neural layer. Meanwhile, the CDiT estimator at each iteration remains trainable and is supervised by the reconstruction loss.

Since our diffusion estimator adopts the $x$-prediction objective, each iteration is directly supervised by the ground truth channel $\mathbf{H}_0$. The reconstruction loss at the $k$-th iteration is given by
\begin{equation}
    \mathcal{L}^{(k)}=
\left\|
\mathbf{H}_0
-
\widehat{\mathbf{H}}_0^{(k)}
\right\|_2^2 .
\label{eq:eudr_stage_loss}
\end{equation}

The feedback cannot be constructed as a fixed input for a non-iterative architecture, because it is generated from the output of each receiver iteration and evolves throughout the refinement process. The estimator should therefore be trained within the iterative loop so that it can adapt to the iteration-dependent distribution of feedback observations. Thus the overall Diff-Rx training objective is defined as the average reconstruction loss over all iterations:
\begin{equation}
 \mathcal{L}_{\mathrm{Diff-Rx}}(\boldsymbol{\theta})= \mathbb{E}_{\mathbf{H}_0,\boldsymbol{\epsilon},t}
\left[
\frac{1}{K}
\sum_{k=0}^{K-1}
\left\|
\mathbf{H}_0
-
\widehat{\mathbf{H}}_0^{(k)}
\right\|_2^2
\right].
\label{eq:training:eudr_training_loss}
\end{equation}
Correspondingly,  the loss is accumulated over all iterations and the network parameters are updated by taking a gradient descent step on
\begin{equation}
\nabla_{\boldsymbol{\theta}}
\mathcal{L}_{\mathrm{Diff-Rx}}(\boldsymbol{\theta}) .
\label{eq:training:eudr_gradient_update}
\end{equation}
This multi-stage supervision encourages every CDiT iteration to produce a valid channel estimation, rather than only optimizing the final output. As a result, the network can learn a progressive refinement behavior. The corresponding training procedure is summarized in Algorithm~\ref{alg:entropy_unfolding_training}.

\subsection{Inference Method}

\subsubsection{Inference of CDiT}
Since CDiT is trained with the $x$-prediction objective, its output is already an estimate of the target channel. Therefore, CDiT naturally supports the one-step generation. Specifically, at the inference stage, we can start from a Gaussian sample at the terminal diffusion step and directly feed it into CDiT together with the conditions and timestep $T$. The network then produces the channel estimation in a single forward pass.

\begin{algorithm}[t]
\caption{Inference Algorithm of Diff-Rx}
\label{alg:eudr_inference}
\begin{algorithmic}[1]
\REQUIRE Received signal \(\mathbf{Y}\), pilot LS estimate \(\mathbf{H}_{\mathrm{LS}}\), pilot mask \(\mathbf{M}\), SNR \(r\), maximum number of iterations \(K\), constellation \(\mathcal{S}\).
\STATE Sample \(\mathbf{H}_T\sim\mathcal{N}(\mathbf{0},\mathbf{I})\)
\STATE Estimate the amplitude factor $\widehat{a}$ using (\ref{eq:inference:amp_norm}) and obtain $\mathbf{H}_{\mathrm{LS}}\leftarrow\mathbf{H}_{\mathrm{LS}}/\widehat{a}$.
\STATE Initialize \(\mathbf{C}^{(0)}=\mathrm{concat}(\mathbf{H}_{\mathrm{LS}}\otimes\mathbf{M},1-\mathbf{M})\)
\FOR{\(k=0\) to \(K-1\)}
\STATE \(\widehat{\mathbf{H}}_0^{(k)}
=
\mathbf{G}_{\boldsymbol{\theta}}
(\mathbf{H}_T,T,\mathbf{C}^{(k)})\).
\IF{early-exit criterion is satisfied}
\RETURN \(\widehat{\mathbf{H}}_0^{(k)} \times\widehat{a}\) .
\ENDIF
\STATE $\widehat{\mathbf{H}}_0^{(k)} = \widehat{\mathbf{H}}_0^{(k)}\times\widehat{a}$.
\STATE Compute posterior probabilities over \(\mathcal{S}\).
\STATE Compute normalized pseudo-channel.
\STATE Update \(\mathbf{C}^{(k+1)}\).
\ENDFOR
\STATE Obtain $\widehat{\mathbf{x}}$ according to \eqref{eq:detection}.
\RETURN \(\widehat{\mathbf{H}}_0^{(K-1)}\) and $\widehat{\mathbf{x}}$.
\end{algorithmic}
\end{algorithm}

The model can also be used with multi-step sampling if a longer denoising trajectory is desired. In this case, a subsequence of diffusion timesteps is selected from the full training schedule. At each selected timestep, CDiT first predicts the channels, and the corresponding noise estimate can then be recovered from the forward diffusion relation in (\ref{eq:Ht}). This recovered noise estimate is used by a standard DDPM scheduler to update the intermediate noisy sample \cite{song2020denoising}. Thus, the proposed CDiT model supports both one-step low-latency generation and conventional multi-step diffusion sampling. In this work, we adopt the one-step generation for efficient receiver inference. 

\subsubsection{Inference of Diff-Rx}

At each iteration, CDiT performs one-step channel generation under the current condition, following the prediction form in (\ref{eq:training:eudr_stage_prediction}). The resulting channel estimate is then used to update the data-symbol posterior, pseudo-channel, and the next condition according to the condition update equations in (\ref{eq:iterative:condition_update_post})--(\ref{eq:iterative:next_cond}).

Although Diff-Rx is trained with a maximum number of iterations $K$, the inference process does not necessarily need to run all $K$ iterations. Since each iteration directly outputs a valid channel estimation, the receiver can terminate early and use the intermediate estimate and detection results as the final output when a strict latency constraint is required. This early-exit property provides a flexible trade-off between performance and inference speed. 

\section{Numerical Results}
This section evaluates CDiT and Diff-Rx in terms of the accuracy of channel estimation and that of data detection under different levels of noise and pilot patterns. We also evaluate the computational complexity of the proposed method.

\subsection{Experiment Details\label{exp:experiment detials}}

\subsubsection{Data generation}
For our experiments, we use Sionna \cite{sionna} to generate the training, validation, and testing datasets. To evaluate the proposed method under both standardized statistical channels and site-specific geometric channels, we consider two types of channel datasets. We generate 24,000 statistical 3GPP urban microcell (UMi) samples and 8,000 site-specific ray-tracing (RT) samples from the Munich Frauenkirche scene.
Both datasets use a 3.5~GHz uplink, an 8-element dual-polarized linear antenna array (ULA), and 512 subcarriers. The subcarrier spacing of UMi channels is 60~kHz and that of RT channels is 120~kHz. For the RT dataset, UEs are randomly distributed at distances ranging from 1 to 400 m and assigned speeds from 0 to 10 m/s, and one of 14 OFDM symbols is selected for each channel sample after retaining the 48 strongest paths. Thus the shape of both types of channels is $16 \times 512$. The relevant parameters of the datasets are summarized in Table \ref{tab:sionna_parameter}.

\begin{table}[!t]
  \centering
  \footnotesize
  \caption{ PARAMETER SETTINGS FOR SIONNA DATASETS}
  \begin{tabularx}{\linewidth}{
    >{\raggedright\arraybackslash\hsize=0.6\hsize}X  
    >{\raggedright\arraybackslash\hsize=0.4\hsize}X  
  }
  \toprule  
    \textbf{Parameters} & \textbf{Value} \\
    \midrule  
    Statistical channel model & UMi \\
    RT scene & Munich \\
    Center frequency & 3.5 GHz \\
    Subcarrier spacing & 60 kHz for UMi \\
                    & 120 kHz for RT\\
    Antenna array form & ULA \\
    Polarization of antennas & Cross-polarization \\
    $N_{\mathrm{r}}$ & 8 \\
    $N_\mathrm{f}$ & 512 \\
    Maximum number of paths & 48 \\
    Maximum number of ray interactions & 5 \\
    Position of BS & [-15, -165, 95] m \\
    Distance between UEs and the BS & 1 m $\sim$ 400 m \\
    Speed of UEs & 0 $\sim$ 10 m/s \\
    \bottomrule  
  \end{tabularx}
  \label{tab:sionna_parameter}
\end{table}

\begin{table}[!t]
  \centering
  \footnotesize
  \caption{PARAMETER SETTINGS FOR TRAINING} 
  \begin{tabularx}{\linewidth}{
    >{\raggedright\arraybackslash\hsize=0.35\hsize}X  
    >{\raggedright\arraybackslash\hsize=0.65\hsize}X  
  }
    \toprule
    Parameters & Value \\
    \midrule
    Batch size & 64 \\
    Optimizer & AdamW \\
    Base learning rate & 3e-4 \\
    LR schedule & Cosine annealing with linear warmup \\
    Pre-training epochs & 2000 \\
    Post-training epochs  & 900 \\
    Timestep $T$ & 1000 \\
    Pilot interval $P$ & $\{1, 2, 4, 8, 16, 32\}$ \\
    Patch size ($p_\mathrm{f},p_\mathrm{r}$) & (8, 8)\\
    SNR  & -10$\sim$30 dB \\
    Structure for CDiT &  $K_{\mathrm{layer}}=3$, $d=192$\\
    Iterations for Diff-Rx & $K=5$ \\
    Parameters for diffusion & $\beta_1=0.0001$, $\beta_{T}=0.02$ \\ 
    \bottomrule
  \end{tabularx}
  \label{tab:training_params}
  \vspace{-0.3cm}
\end{table}

\subsubsection{Performance metrics}
We use normalized MSE (NMSE) \cite{8322184} between the estimated channel and true channel and bit error rate (BER) as the performance indexes, which are defined as follows:
\begin{equation}
\text{NMSE} = \mathbb{E} \left\{ {\|\textbf{H}_0 - \widehat{\textbf{H}}_0\|_{\mathrm{F}}^2}/{\|\textbf{H}_0\|_{\mathrm{F}}^2} \right\}.
\label{eq:nmse}
\end{equation}

\begin{figure*}[t!]
  \centering
  \subcaptionbox{Comparison on UMi.\label{fig:exp:CE:1}}[0.33\linewidth]{%
  
    \centering
    \includegraphics[width=\linewidth]{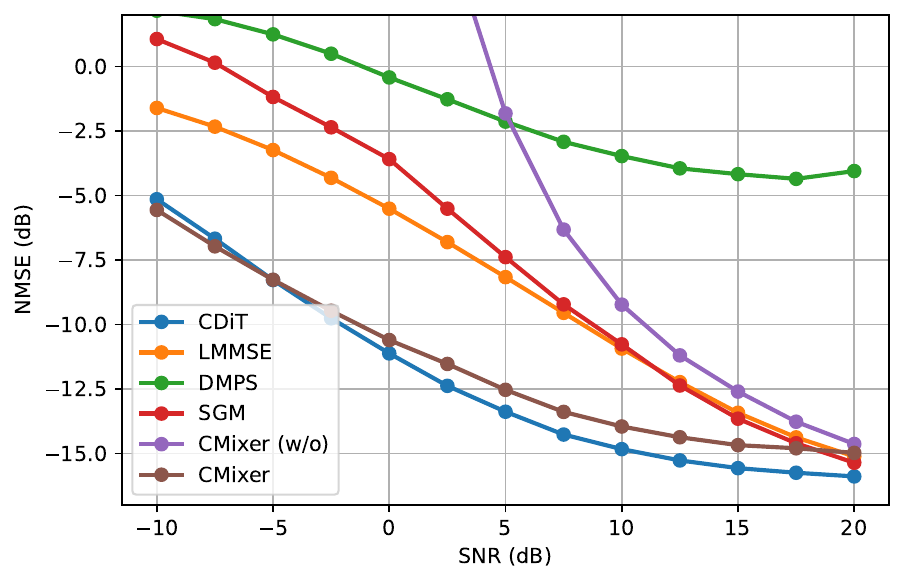}%
  }
\hfill 
  \subcaptionbox{Comparison on RT.\label{fig:exp:CE:2}}[0.33\linewidth]{%
    
    \centering
    \includegraphics[width=\linewidth]{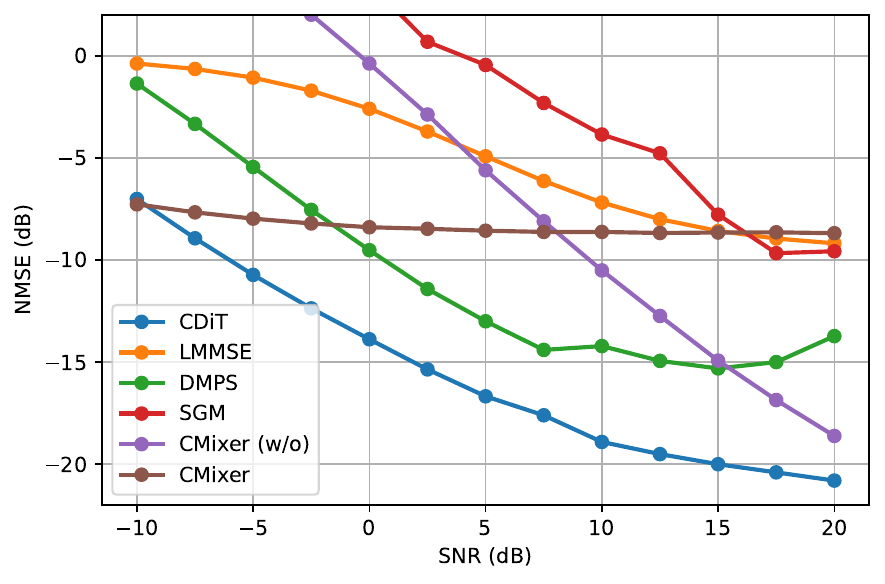}%
  }%
  \hfill 
  \subcaptionbox{Comparison on RT with different pilot densities.\label{fig:exp:CE:3}}[0.33\linewidth]{%
    \centering
    \includegraphics[width=\linewidth]{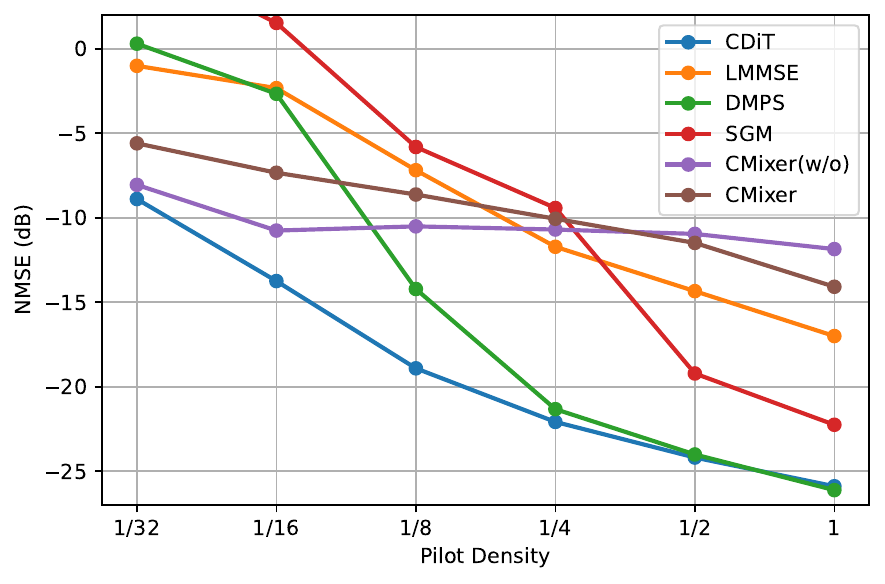}%
  }%
    \caption{Comparison of channel estimation only based on pilot observation on UMi and RT channels with 4-QAM. (a) UMi channels with $P=16$. (b) RT channels with $P=8$. (c) RT channels under different values of $P$ with $\mathrm{SNR} = 10~\mathrm{dB}$.
}
  \label{fig:exp:CE}
\end{figure*}

\begin{figure*}[t!]
  \centering

  \subcaptionbox{Comparison under different SNR. \label{fig:exp:udce:nmse_compare:1}}[0.33\linewidth]{%
    \centering
    \includegraphics[width=\linewidth]{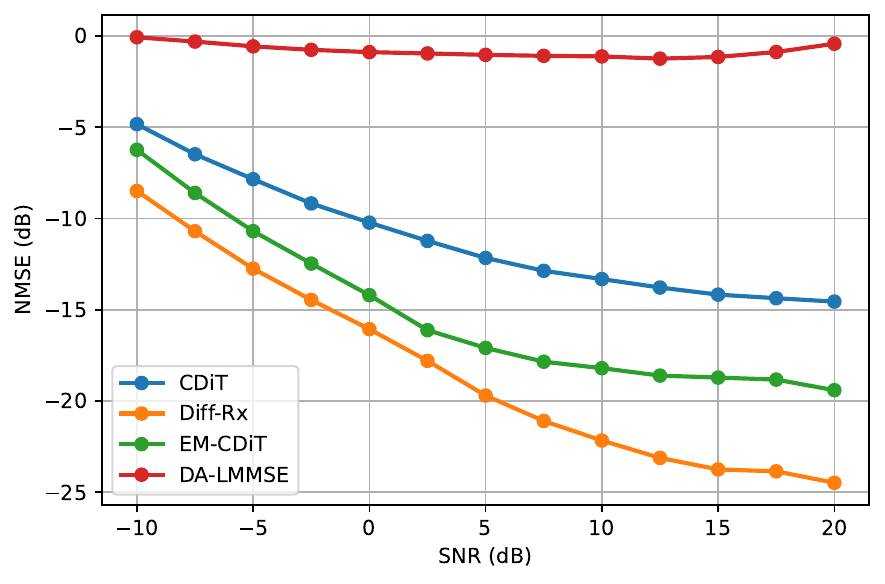}%
  }%
  \hfill 
  \subcaptionbox{Comparison under different pilot density. \label{fig:exp:udce:nmse_compare:2}}[0.33\linewidth]{%
    \centering
    \includegraphics[width=\linewidth]{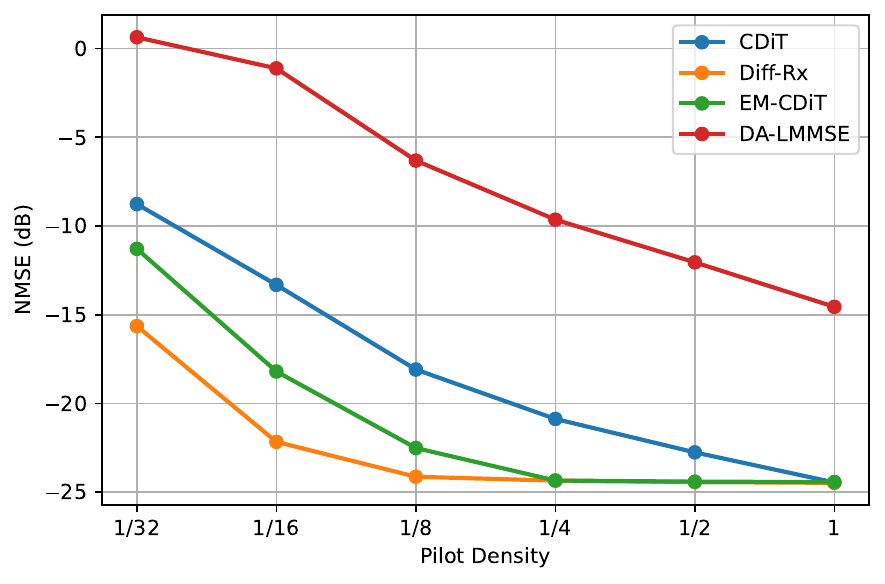}%
  }%
  \hfill 
  \subcaptionbox{Comparison under different constellations. \label{fig:exp:udce:nmse_compare:3}}[0.33\linewidth]{%
    \centering
    \includegraphics[width=\linewidth]{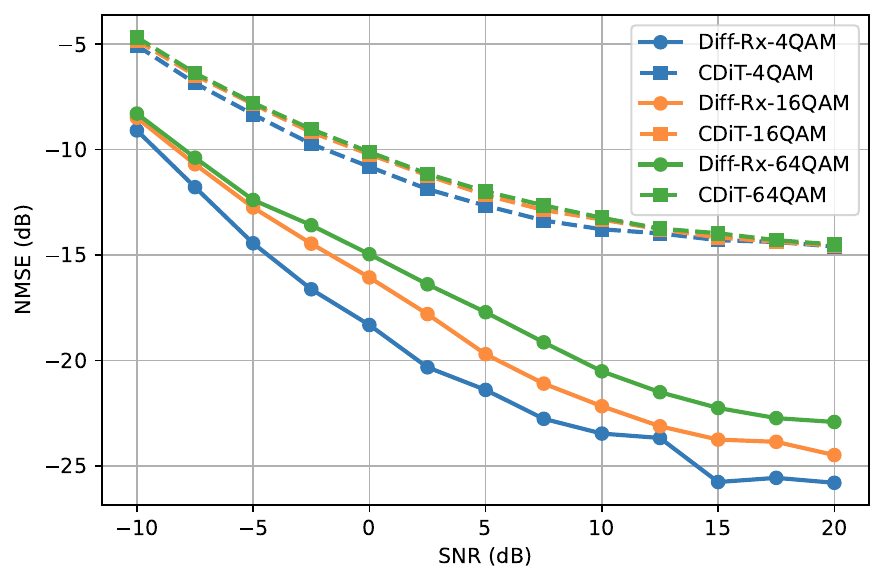}%
  }

  \subcaptionbox{16-QAM and $P=4$.\label{fig:exp:udce:ber:16QAM_P4}}[0.33\linewidth]{%
    \centering
    \includegraphics[width=\linewidth]{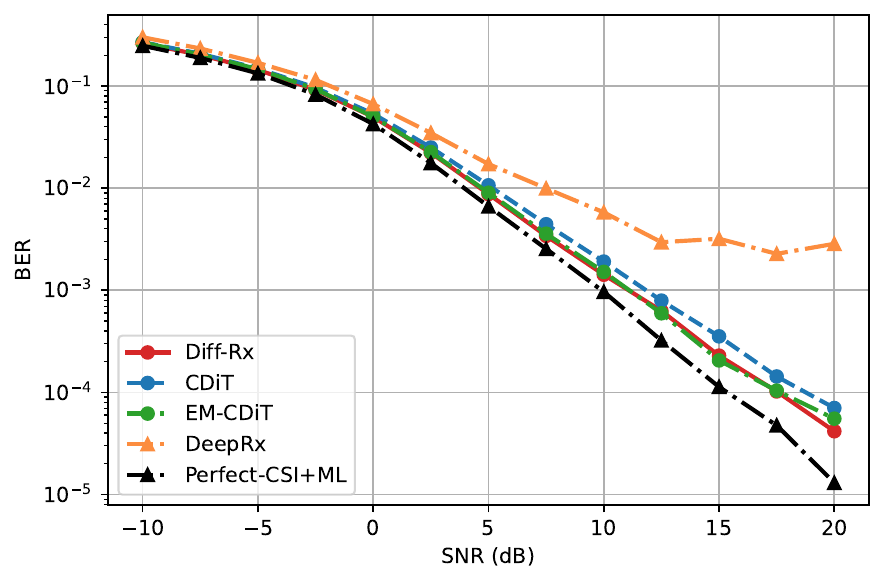}%
  }%
  \hfill 
  \subcaptionbox{16-QAM and $P=16$.\label{fig:exp:udce:ber:16QAM_P16}}[0.33\linewidth]{%
    \centering
    \includegraphics[width=\linewidth]{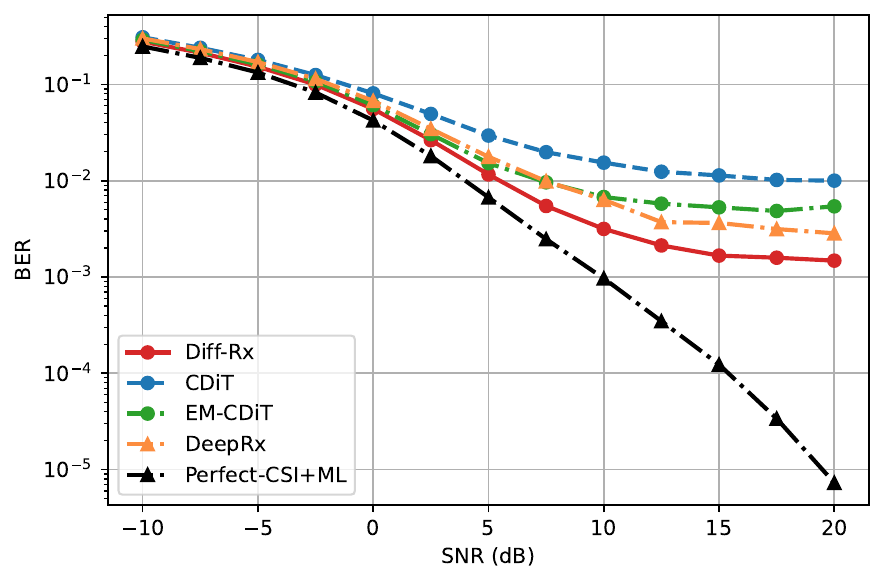}%
  }%
   \hfill 
  \subcaptionbox{64-QAM and $P=16$.\label{fig:exp:udce:ber:64QAM_P16}}[0.33\linewidth]{%
    \centering
    \includegraphics[width=\linewidth]{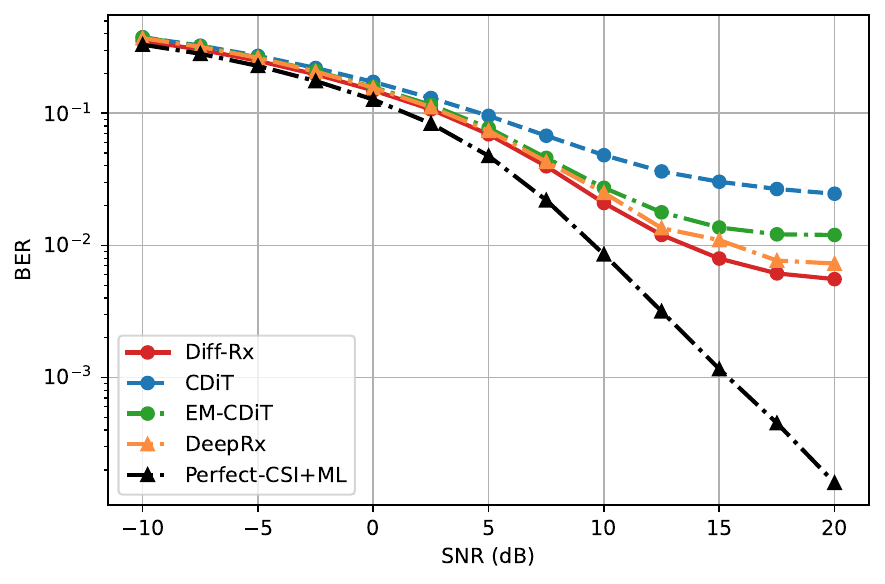}%
  }
\hfill 

  \caption{Performance of Diff-Rx on RT channels. (a)-(c) compare the channel estimation performance of the different methods and (d)-(f) compare the detection performance of the different methods. The QAM symbols are mapped using Gray code in the experiments. In (a) and (b), the data symbols are selected from 16-QAM. The thresholds of EM-CDiT and DA-LMMSE are 0.5. In (a) and (c), we set $P = 16$.}
  \label{fig:exp:udce:receiver}
  \vspace{-0.4cm}
\end{figure*}

\subsubsection{Baselines and training settings}
During the training process, the SNR of the received signals is randomly distributed within the range of -10 to 30 dB. The pilot interval $P$ is randomly selected from the set $\{1, 2, 4, 8, 16, 32\}$. We train on two different datasets with the same training parameters.  The relevant parameters of the training are detailed in Table \ref{tab:training_params}. The following baseline methods are compared:
\begin{itemize}
    \item \textbf{LMMSE}: The algorithm computes the channel estimate and the corresponding error variance for each element of the OFDM resource grid using a linear estimator based on the sample covariance calculated from all training samples. Frequency-domain interpolation is then performed. 
    \item \textbf{CMixer}: This model uses two complex-domain MLP modules to separately learn spatial and frequency characteristics. We set the number of layers in CMixer to 4 to ensure that the FLOPs of CMixer remain comparable to that of CDiT. The training details and other parameters follow \cite{10445518}. 
    \item \textbf{Score-based generative model (SGM)} \cite{9957135}: The SGM-based estimator learns a score-based generative prior and performs estimation using annealed Langevin dynamics. 
    \item \textbf{Diffusion model posterior sampling (DMPS)} \cite{10930691}: The DMPS-based estimator learns a prior for MIMO channels and performs posterior inference by combining the DM prior score with the measurement likelihood score. We find that the CNN network used in the paper does not yield desirable performance in our OFDM channel estimation task. Therefore, we replace the CNN network with a DiT model. In addition, we adjust the scaling coefficient for different pilot densities so that the network could achieve the best estimation performance. 
    \item \textbf{CDiT}: This baseline does not perform post-training and directly performs channel estimation based on the pilot observations.
    \item \textbf{DA-LMMSE (data-aided LMMSE)} \cite{khan2023data}: This conventional approach evaluates symbol reliability based on the Euclidean distance between equalized constellation points and ideal constellation symbols and then performs LMMSE channel estimation.  
    \item \textbf{EM-CDiT}: This baseline embeds CDiT into a conventional EM-style iterative procedure. After each channel estimation step, data detection is performed using maximum a posteriori (MAP) detection, and reliable data symbols are selected as virtual pilots according to a posterior probability threshold.
    \item \textbf{Perfect CSI+ML}: This baseline has perfect CSI and uses maximum likelihood (ML) detection.
    \item \textbf{DeepRx} \cite{honkala2021deeprx}: A fully convolutional deep-learning receiver that jointly performs channel estimation, equalization, and detection and directly outputs the detection results. 
\end{itemize}

For the considered signal model in (\ref{eq:channel model2}), we vectorize the observation as $\mathbf{y}=\mathbf{A}\mathbf{h}+\mathbf{n}$, with $\mathbf{y}=\mathrm{vec}(\mathbf{Y})$, $\mathbf{h}=\mathrm{vec}(\mathbf{H})$, and $\mathbf{A}=\mathbf{X}^{\mathrm{T}}\odot \mathbf{I}$, where $\odot$ represents the Kronecker product. During posterior inference, an effective measurement matrix is constructed by preserving only the known pilot entries in $\mathbf{X}$ and setting the unknown entries of data symbols to zero, such that the posterior update is guided only by the available pilot observations. In this way, the MIMO channel estimation problem considered in the baseline scheme \cite{9957135, 10930691} is aligned with our OFDM channel estimation setting.

\subsection{Performance of Channel Estimation Based on Pilots}

We first evaluate the performance of the CDiT in both UMi channels and RT channels. The models are trained and tested separately on the two datasets. The results are shown in Fig.~\ref{fig:exp:CE}. The proposed CDiT model demonstrates strong pilot adaptability and noise robustness. Under different SNR levels and pilot densities, CDiT consistently outperforms the baseline methods. Compared with the UMi dataset, the RT dataset has a larger subcarrier spacing and fewer training samples. It also contains more diverse multipath types and a larger number of propagation paths, making training and generalization more challenging. Nevertheless, CDiT still outperforms the baseline methods on this more difficult dataset. SGM and DMPS can achieve good performance when the pilot density is high. However, when the pilot density is lower than 1/8, their performance degrades significantly. This indicates that both methods are highly sensitive to the completeness of pilot observations and have difficulty maintaining reliable channel reconstruction performance with sparse pilots. Moreover, the performance of these two baselines is sensitive to hyperparameter settings. Under different SNR levels, pilot densities, and datasets, the scaling coefficient must be selected carefully and differently to achieve optimal performance, and even slight mismatches in this parameter can lead to noticeable performance degradation. The curves of SGM and DMPS in Fig.~\ref{fig:exp:CE:1} and~\ref{fig:exp:CE:2} exhibit an upward trend under high-SNR conditions because we select a near-optimal scaling coefficient for each pilot density, while keeping it fixed across different SNR levels without further tuning for different SNR. Meanwhile, CDiT does not introduce any additional coefficient, thereby avoiding performance degradation caused by inappropriate parameter selection. CMixer uses a fixed pilot pattern and shows limited robustness to noise variations. In Fig.~\ref{fig:exp:CE}, CMixer (w/o) takes the noise-free masked channel $\textbf{H}\otimes\textbf{M}$ as input and degrades at low SNR, whereas the default CMixer uses the raw noisy estimate $\widetilde{\textbf{H}}$, improving low-SNR robustness at the cost of high-SNR accuracy.

\begin{figure}[t!]
    \centering

    \begin{subfigure}[b]{1\linewidth}
        \centering
        \includegraphics[width=0.8\linewidth, keepaspectratio]{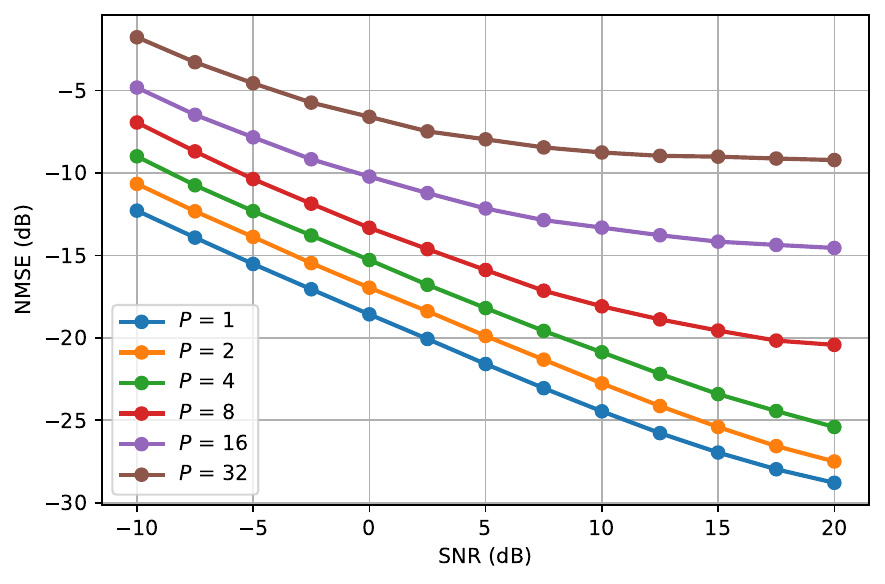}
        \caption{Performance of CDiT in different SNR and $P$.}
        \label{fig:exp:all:cdit}
    \end{subfigure}

    \begin{subfigure}[b]{1\linewidth}
        \centering
        \includegraphics[width=0.8\linewidth, keepaspectratio]{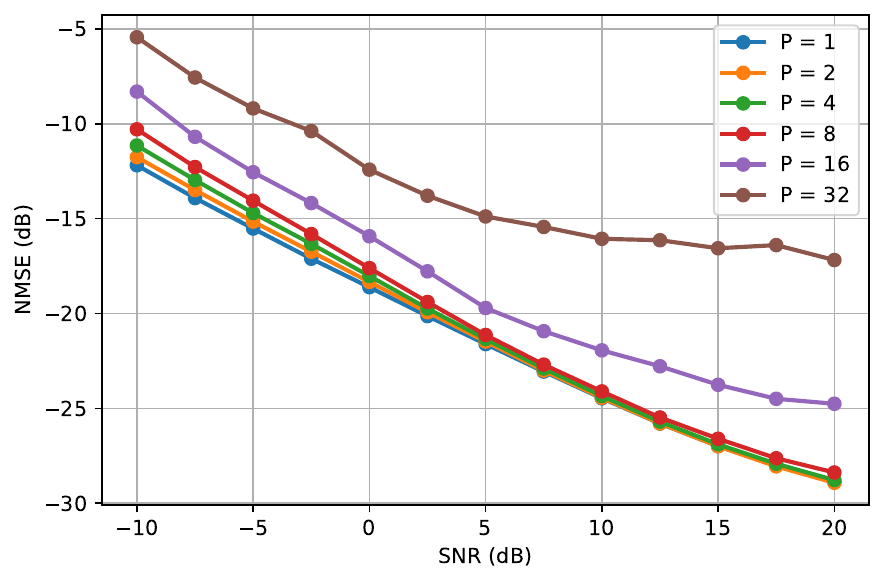}
        \caption{Performance of Diff-Rx in different SNR and $P$.}
        \label{fig:exp:all:udce}
    \end{subfigure}
    
    \caption{Comparison of channel estimation between CDiT and Diff-Rx.}
    \label{fig:exp:cdit_udce}
    \vspace{-0.5cm}
\end{figure}

\subsection{Performance of Diff-Rx}
We evaluate the performance of channel estimation and data detection of Diff-Rx on RT channels and the results are shown in Fig.~\ref{fig:exp:udce:receiver}. Compared with the baselines in Fig.~\ref{fig:exp:udce:nmse_compare:1} and \ref{fig:exp:udce:nmse_compare:2}, Diff-Rx further improves performance by jointly learning data-symbol utilization and channel estimation within a unified trainable framework, enabling more effective exploitation of both pilot and data observations. In contrast, methods that separate reliable-symbol selection from channel estimation may not fully utilize the channel information from data observations. For DA-LMMSE, the initial channel estimation is inaccurate under sparse pilot observations. At high SNR, wrong virtual pilots are assigned very small observation noise variance and are thus overtrusted by the LMMSE update, resulting in performance degradation as SNR increases.

To evaluate the generalizability of Diff-Rx across different modulations, we use pilot symbols from the 16-QAM constellation for the pretrained CDiT and employ 4-QAM, 16-QAM, and 64-QAM data symbols during post-training. The results are shown in Fig.~\ref{fig:exp:udce:nmse_compare:3}, with the remaining training parameters kept the same as those in Table \ref{tab:training_params}. Since the pretrained CDiT relies only on pilot symbols for channel estimation, changing the data-symbol modulation has little impact on its performance, resulting in nearly identical curves. In contrast, the performance of Diff-Rx varies across different modulations while consistently outperforming the CDiT only based on pilot observations. Its performance decreases as the modulation order increases because the reduced constellation spacing makes symbol detection and channel information extraction more difficult. Nevertheless, Diff-Rx still maintains clear performance gains. This confirms that the improvement of Diff-Rx arises from the joint use of pilot and data symbols for channel estimation.

\begin{table}[t!]
    \centering
    \footnotesize
    \renewcommand{\thetable}{\Roman{table}}
    \newcolumntype{Y}{>{\centering\arraybackslash}X}
    \newcolumntype{C}[1]{>{\centering\arraybackslash}m{#1}}
    \caption{EFFECT OF DATA SYMBOL SELECTION STRATEGIES ON NMSE PERFORMANCE.}
    \label{exp:tab:udce:threshold}
    \renewcommand{\arraystretch}{1.25}
    \setlength{\tabcolsep}{3pt}

    \begin{tabularx}{\columnwidth}{@{}
        C{0.85cm}
        *{4}{Y}
        C{0.75cm}
    @{}}
        \toprule
        \makecell[c]{SNR\\(dB)}
        & \makecell[c]{Diff-Rx\\(proposed)}
        & \makecell[c]{Diff-Rx-hard\\($m=0$)}
        & \makecell[c]{Diff-Rx-hard\\($m=0.5$)}
        & \makecell[c]{Diff-Rx-hard\\($m=1$)}
        & CDiT \\
        \midrule

        -10
        & $\mathbf{-8.31}$
        & -7.96
        & -7.16
        & -5.57
        & -4.86 \\

        -5
        & $\mathbf{-12.56}$
        & -11.88
        & -11.93
        & -8.49
        & -7.89 \\

        0
        & $\mathbf{-15.93}$
        & -15.57
        & -15.59
        & -10.83
        & -10.17 \\

        10
        & -21.94
        & -21.93
        & $\mathbf{-21.95}$
        & -14.45
        & -13.46 \\

        20
        & $\mathbf{-24.76}$
        & -24.43
        & -24.18
        & -15.87
        & -14.44 \\

        \bottomrule
    \end{tabularx}
\end{table}
\begin{table}[t!]
    \centering
    \footnotesize
    \renewcommand{\thetable}{\Roman{table}}

    \newcolumntype{Y}{>{\centering\arraybackslash}X}
    \newcolumntype{C}[1]{>{\centering\arraybackslash}m{#1}}

    \caption{SNR MISMATCH ON NMSE PERFORMANCE.}
    \label{exp:tab:udce:snrmismatch}

    \renewcommand{\arraystretch}{1.15}
    \setlength{\tabcolsep}{2.8pt}

    \newcommand{\datarowstrut}{\rule[-1.1ex]{0pt}{3.5ex}}

    \begin{tabularx}{\columnwidth}{@{}
        C{0.8cm}
        *{7}{Y}
    @{}}
        \toprule

        \multirow[c]{2}{*}[-0.8ex]{%
            \makecell[c]{Assumed\\SNR\\(dB)}%
        }
        & \multicolumn{7}{c}{Actual SNR Mismatch (dB)} \\
        \cmidrule(lr){2-8}

        & $-3$ & $-2$ & $-1$ & $0$ & $1$ & $2$ & $3$ \\
        \midrule

        \datarowstrut $-10$
        & $-6.65$ & $-7.49$ & $-8.16$ & $-8.31$
        & $-8.17$ & $-7.14$ & $-5.59$ \\

        \datarowstrut $0$
        & $-11.27$ & $-13.28$ & $-15.20$ & $-15.93$
        & $-15.10$ & $-13.05$ & $-10.95$ \\

        \datarowstrut $10$
        & $-19.59$ & $-21.37$ & $-22.02$ & $-21.94$
        & $-21.64$ & $-20.95$ & $-20.94$ \\

        \datarowstrut $20$
        & $-24.62$ & $-25.10$ & $-24.67$ & $-24.76$
        & $-25.14$ & $-24.01$ & $-24.16$ \\

        \bottomrule
    \end{tabularx}
    \vspace{-0.3cm}
\end{table}

Figs.~\ref{fig:exp:udce:ber:16QAM_P4}--\ref{fig:exp:udce:ber:64QAM_P16} compare the BER performance under different modulations and pilot densities. Diff-Rx consistently outperforms the baseline methods across all considered settings. Unlike DeepRx, Diff-Rx explicitly alternates between channel estimation and symbol posterior inference, thereby preserving the receiver structure while reducing the burden of learning the entire process of the receiver. Compared with CDiT and EM-CDiT, Diff-Rx can obtain more accurate CSI and consequently improve detection performance. As the pilot density decreases, the increasingly sparse pilot observations limit the channel estimation accuracy, making the CSI errors the primary source of the remaining performance gap to the ideal perfect CSI benchmark.

We compare the channel estimation performance of CDiT and Diff-Rx under different SNRs and pilot densities in Fig.~\ref{fig:exp:cdit_udce}. After post-training, Diff-Rx achieves nearly identical NMSE when $P \leq 8$, approaching the full-pilot performance, and still provides substantial gains under sparser pilot configurations. The superiority of Diff-Rx over the channel estimation based on pilots confirms that data symbols provide useful auxiliary channel information. 

\begin{figure}
    \centering
    \includegraphics[width=0.8\linewidth]{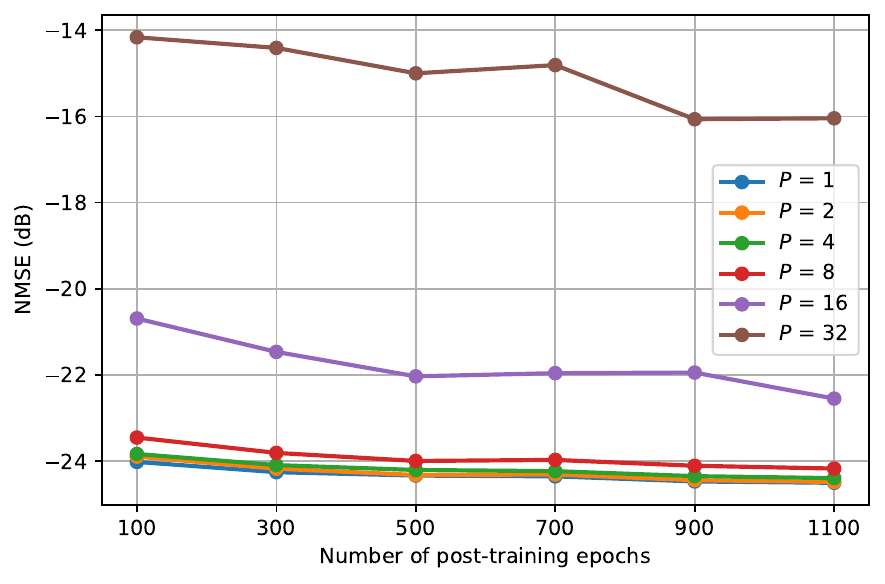}
    \caption{Performance of channel estimation under different post-training epochs with $K=5$.}
    \label{fig:exp:duce:epochs:nmse}
    \vspace{-0.4cm}
\end{figure}

\begin{figure}[t!]
    \centering

    \begin{subfigure}[b]{1\linewidth}
        \centering
        \includegraphics[width=0.8\linewidth, keepaspectratio]{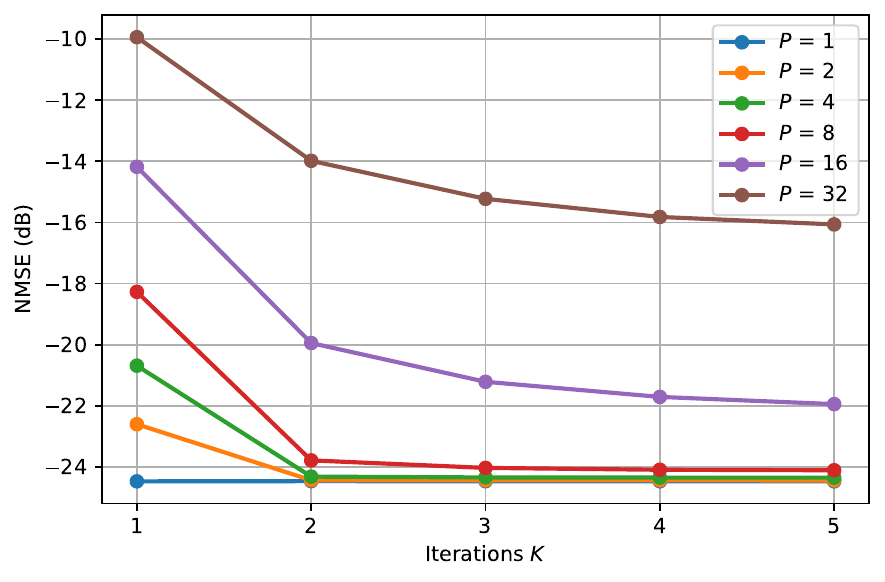}
        \caption{Performance of channel estimation.}
        \label{fig:exp:duce:iter:nmse}
    \end{subfigure}

    \begin{subfigure}[b]{1\linewidth}
        \centering
        \includegraphics[width=0.8\linewidth, keepaspectratio]{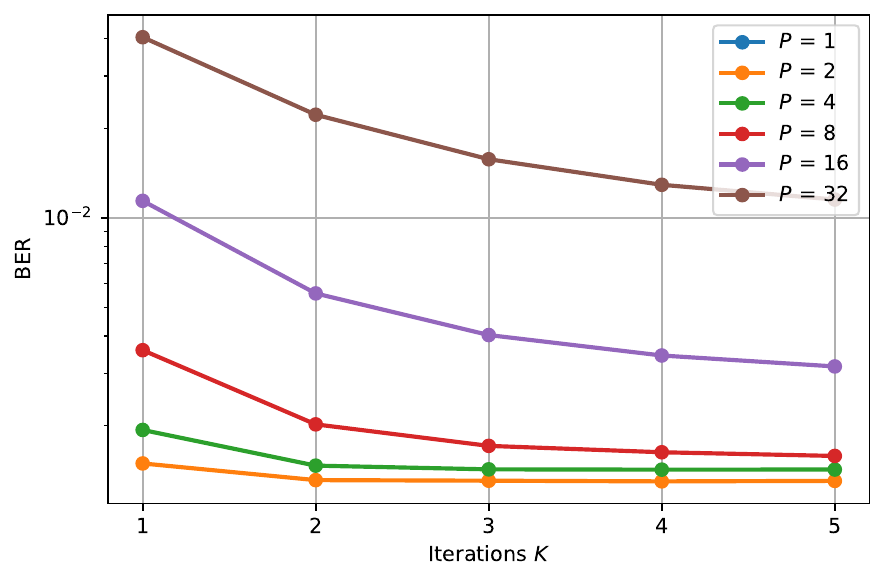}
        \caption{Performance of detection.}
        \label{fig:exp:duce:iter:ber}
    \end{subfigure}

    \caption{Effects of iterations on the performance of channel estimation and detection at $\text{SNR}=10~\text{dB}$.}
    \label{fig:exp:udce:iter}
    \vspace{-0.3cm}
\end{figure}

Fig.~\ref{fig:exp:udce:iter} evaluates the impact of the number of Diff-Rx iterations on channel estimation. The case $K=1$ corresponds to channel estimation using only the pilot positions. When $P \leq 8$, Diff-Rx almost reaches its best NMSE performance after only two iterations. Under sparser-pilot settings, the performance also nearly converges after three iterations, indicating that Diff-Rx can achieve significant channel estimation gains with a very small iterative cost. 
The BER results in Fig.~\ref{fig:exp:duce:iter:ber} show a trend consistent with the NMSE results. The BER decreases rapidly within the first few iterations and then gradually converges. This further verifies that Diff-Rx can effectively couple channel estimation and symbol detection, improving both NMSE and BER performance with low iterative complexity. 
As shown in Fig.~\ref{fig:exp:duce:epochs:nmse}, when the pilot density is high, the model converges with only a small number of post-training epochs. Even under low pilot density, the required number of post-training epochs remains limited and is much smaller than that used for pretraining. Thus, Diff-Rx supports the low-cost adaptation to different constellations.

Table \ref{exp:tab:udce:threshold}  compares the proposed soft decision with hard-decision strategies denoted as Diff-Rx-hard under $P=16$. For the hard-decision method, the threshold is denoted as $m$. A data symbol is assigned to the constellation point with the maximum posterior probability only when this probability exceeds $m$. Thus, $m=0$ selects all data symbols, whereas $m=1$ selects none and reduces to the channel estimation based on pilots. The optimal hard-decision threshold is related to the SNR and the pilot density, while the proposed soft strategy avoids threshold tuning and achieves consistently superior performance. The gain is more significant at low SNRs, where hard decisions are less reliable, and it becomes smaller at high SNRs as the posterior probabilities become more accurate. Diff-Rx-hard with $m=1$ still outperforms the pretrained CDiT. This improvement mainly comes from different training configurations rather than data symbol selection. Specifically, Diff-Rx-hard uses uniformly spaced pilots whose indices are $P_{\mathrm{start}}+wP$, while the pretrained CDiT is trained with additional randomly inserted pilots for arbitrary pilot patterns. This means that the pretrained CDiT must adapt to a larger number of pilot patterns. Moreover, Diff-Rx-hard is trained for additional epochs during post-training.

Although the proposed model assumes that the SNR is known, it may deviate from the actual value in the real system. We therefore consider SNR mismatches within $\pm 3$~dB. The results are shown in Table \ref{exp:tab:udce:snrmismatch}. In the table, the leftmost column denotes the SNR value assumed and used by the receiver, whereas the 'Actual SNR Mismatch' indicates the deviation of the assumed SNR from the actual SNR. Such mismatches affect both the normalization amplitude in (\ref{eq:eudr_amp_estimation}) and the posterior-probability calculation. The former causes the normalized channel to deviate from the training distribution, whereas the latter introduces errors into the soft decision and pseudo-channel estimates.

Nevertheless, the results show that Diff-Rx maintains stable performance under these mismatches, demonstrating strong robustness to SNR mismatches.

\begingroup
\begin{table}[t!]
    \centering
    \footnotesize
    \renewcommand{\thetable}{\Roman{table}}
    \caption{PERFORMANCE AND COMPLEXITY COMPARISON OF DIFFERENT METHODS.}
    \label{exp:tab:model_size}

    \renewcommand{\arraystretch}{1.45}
    \setlength{\tabcolsep}{3pt}

    \renewcommand{\tabularxcolumn}[1]{m{#1}}

    \begin{tabularx}{\columnwidth}{%
        >{\centering\arraybackslash}m{1.35cm}
        | >{\centering\arraybackslash}X
        | >{\centering\arraybackslash}X
        | >{\centering\arraybackslash}X
    }
        \hline

        \textbf{Model}
        & \textbf{Number of\newline parameters}
        & \textbf{FLOPs per iteration}
        & \textbf{Inference steps}
        \\
        \hline

        CDiT
        & 3.02~M
        & 478.40~M
        & 1
        \\
        \hline

        Diff-Rx
        & 3.02~M
        & 478.40~M
        & 5
        \\
        \hline

        DMPS
        & 3.89~M
        & 635.10~M
        & 100
        \\
        \hline

        SGM
        & 3.31~M
        & 7.41~G
        & 2311
        \\
        \hline

        CMixer
        & 17.93~M
        & 591.46~M
        & 1
        \\
        \hline

        DeepRx
        & 0.66~M
        & 10.90~G
        & 1
        \\
        \hline
    \end{tabularx}

    \vspace{-0.3cm}
\end{table}
\endgroup

\subsection{Complexity of Methods}
The model sizes and FLOP counts of the different methods are shown in Table~\ref{exp:tab:model_size}. Compared with the baseline methods, the proposed CDiT network maintains a controllable model complexity in terms of both the number of parameters and computational cost, while achieving superior performance. Moreover, by adopting a one-step generation strategy, the proposed CDiT network substantially improves inference efficiency and significantly reduces inference latency compared with SGM and DMPS, which rely on a large number of iterative inference steps.

\section{Conclusion}


In this paper, we proposed Diff-Rx, an incrementally conditioned diffusion-based AI receiver for MIMO-OFDM systems. The key idea is to make receiver responsive to a conditioning state that evolves throughout iterations. By post-training the channel estimator which only trained based on pilots within the iterative loop, Diff-Rx adapts to data-derived observations whose reliability evolves across iterations and effectively exploits their correlations with pilot observations. Together with threshold-free soft decision feedback and one-step generation for low-latency inference, the proposed framework provides an efficient and interpretable approach to joint channel estimation and data detection.

Future work will further explore methods to improve the scenario generalization and adaptivity of the proposed framework, such as incorporating neighborhood information \cite{10845822}, digital-twin models, and reinforcement-learning architectures that enable interaction and feedback between the network and wireless environments. Though Diff-Rx can be extended to multi-stream systems, the resulting inter-stream interference may make posterior inference over transmitted symbols and data-aided conditioning updates more challenging.

\bibliographystyle{IEEEtran}
\bibliography{myref}
\end{document}